\documentclass[12pt]{article}

\usepackage[utf8]{inputenc}

\usepackage{authblk}
\usepackage{amsmath}
\usepackage{graphicx}
\graphicspath{{./images/} }
\usepackage{physics}
\usepackage{xcolor}
\usepackage[english]{babel}
\usepackage{makecell}
\usepackage{float}
\usepackage{geometry}
\usepackage{csquotes}

\usepackage{hyperref}
\usepackage{caption}
\hypersetup{
    colorlinks,
    citecolor=black,
    filecolor=black,
    linkcolor=black,
    urlcolor=black
}

\usepackage[sorting = none, backend = biber, style=numeric-comp, doi = true]{biblatex}
\renewbibmacro{in:}{%
  \ifentrytype{article}{}{\printtext{\bibstring{in}\intitlepunct}}}
\DeclareFieldFormat[article]{citetitle}{#1}
\DeclareFieldFormat[article]{title}{#1}

\title{High precision differential clock comparisons with a multiplexed optical lattice clock}

\author{Xin Zheng$^{1}$, Jonathan Dolde$^{1}$, Varun Lochab$^{1}$, Brett Merriman$^{1}$, Haoran Li$^{1}$, Shimon Kolkowitz$^{1,}$\footnote{To whom correspondence should be addressed. E-mail: kolkowitz@wisc.edu}}
\affil{\textit{\normalsize${}^{1}$University of Wisconsin-Madison}}

\date{\today}

\newgeometry{vmargin={15mm}, hmargin={20mm,25mm}} % set the margins

\begin{document}

\renewcommand{\abstractname}{\vspace{-\baselineskip}} % remove Abstract title

\maketitle

\begin{abstract}
    \normalsize
    \textbf{Rapid progress in the precision and accuracy of optical atomic clocks over the last decade has advanced the frontiers of timekeeping, metrology, and quantum science~\cite{ludlow_optical_2015, campbell_fermi_degenerate_2017, mcgrew_atomic_2018}.
    However, the stabilities of most optical clocks remain limited by the local oscillator rather than the atoms themselves, leaving room for further progress~\cite{schioppo_ultrastable_2017, oelker_demonstration_2019}.
    Here we implement a ``multiplexed'' one-dimensional optical lattice clock,
    in which spatially-resolved, movable ensembles of ultra-cold strontium atoms are trapped in the same optical lattice,
    interrogated simultaneously by a shared clock laser,
    and read-out in parallel.
    By performing synchronized Ramsey interrogations of ensemble pairs
    we observe atom-atom coherence times up to 26 seconds,
    a 270-fold improvement over the atom-laser coherence time,
    demonstrate a relative stability of $9.7(4)\times10^{-18}/\sqrt{\tau}$ (where $\tau$ is the averaging time in seconds),
    and reach a fractional uncertainty of $8.9(3)\times 10^{-20}$ after 3.3 hours of averaging. 
    These results demonstrate that applications requiring ultra-high-precision comparisons between optical atomic clocks need not be limited by the stability of the local oscillator.
    With multiple ensemble pairs, we realize a miniaturized clock network consisting of 6 atom ensembles,
    resulting in 15 unique pairwise clock comparisons with relative stabilities below $3\times10^{-17}/\sqrt{\tau}$.
    Finally, we demonstrate the capability to simultaneously load spatially-resolved, heterogeneous ensemble pairs of all four stable isotopes of strontium in a lattice. 
    The unique capabilities offered by this platform pave the way for future studies of precision isotope shift measurements, 
    spatially resolved characterization of limiting clock systematics,
    development of clock-based gravitational wave and dark matter detectors~\cite{kolkowitz_gravitational_2016, derevianko_hunting_2014, wcislo_dark_2018, kennedy_precision_2020},
    and novel tests of relativity including measurements of the gravitational redshift at sub-centimeter scales~\cite{chou_optical_2010, takano_geopotential_2016, grotti_geodesy_2018, takamoto_test_2020}}
\end{abstract}

Neutral atom optical lattice clocks (OLCs) have recently reached stability and accuracy at the $10^{-18}$ level~\cite{ludlow_optical_2015, hinkley_atomic_2013, mcgrew_atomic_2018, campbell_fermi_degenerate_2017, schioppo_ultrastable_2017,marti_imaging_2018,bothwell_jila_2019, oelker_demonstration_2019} largely due to the ultra-narrow linewidths ($\sim1$ mHz) of optical frequency ($\sim400$ THz) forbidden clock transitions in alkaline-earth(-like) atoms.
This performance enables novel clock applications such as relativistic geodesy,
searches for dark matter,
gravitational wave detection,
and tests of fundamental physics~\cite{chou_optical_2010, arvanitaki_searching_2015, derevianko_hunting_2014, kolkowitz_gravitational_2016, kolkowitz_spinorbit_coupled_2017, grotti_geodesy_2018, safronova_search_2018, wcislo_dark_2018, takamoto_test_2020, beloy_frequency_2021, kennedy_precision_2020,leopardi_measurement_2021, burt_demonstration_2021, takano_geopotential_2016}.

Many emerging clock applications rely on differential comparisons between two or more optical clocks, 
rather than on the determination of absolute frequencies.
For atoms in unentangled states,
the stability of such clock comparisons is fundamentally limited by the quantum projection noise (QPN)~\cite{itano_quantum_1993}.
For Ramsey spectroscopy,
the QPN limit for the statistical fractional frequency uncertainty in clock comparison is given by
\begin{equation}
    \sigma_{\textrm{QPN}}(\tau) = \frac{\sqrt{2}}{2\pi\nu CT}\sqrt{\frac{T + T_{\textrm{d}}}{N \tau}},
\label{QPN_eq}
\end{equation}
where $\nu$ is the transition frequency,
$T$ is the interrogation time,
$T_{\textrm{d}}$ is the dead time between experiment cycle,
$\tau$ is the averaging time,
$N$ is the atom number per clock per measurement,
$C$ is the contrast of Ramsey fringes,
and the factor of $\sqrt{2}$ assumes equal contribution from each clock.
Eq.~\ref{QPN_eq} implies that the stability can be improved with greater atom numbers and longer coherence times.
However,
frequency noise in the clock lasers used to interrogate the atoms results in reduced atom-laser coherence times,
and also prevents the clock stability from reaching the QPN limit for larger atom numbers due to the Dick effect~\cite{dick_g_j_local_1987,dick_g_j_local_1990, bloom_optical_2014},
an aliasing of frequency noise from the non-continuous laser interrogation.
This motivates the use of simultaneous differential comparisons \cite{schioppo_ultrastable_2017, kim_2021_optical}, also known as correlated noise spectroscopy \cite{nicholson_comparison_2012}, for applications involving clock comparisons \cite{takano_geopotential_2016, kolkowitz_gravitational_2016}.
Common-mode rejection of Dick noise and 10-second-scale atom-atom coherence times well beyond that of the clock laser have recently been demonstrated between two independent ion-clocks~\cite{clements_lifetime_limited_2020},
between sub-ensembles in a three-dimensional Fermi-degenerate OLC~\cite{campbell_fermi_degenerate_2017},
and between sub-ensembles in a tweezer-array clock~\cite{young_half_minute_scale_2020}. In each of these cases the atoms are individually and tightly confined,
suggesting that strong confinement and a lack of atom-atom interactions may be necessary ingredients to achieve such long coherent interrogation times. 
Furthermore, the best differential stabilities observed thus far, in the range of $3\times10^{-17}/\sqrt{\tau}$ \cite{oelker_demonstration_2019, campbell_fermi_degenerate_2017,marti_imaging_2018} have made use of an 8 mHz linewidth clock laser with a stability of $4\times10^{-17}$ at 1 s \cite{matei_15text_2017}, 
suggesting that even in simultaneous differential comparisons clock laser coherence could still play a role in limiting the achievable stability.

Here we introduce and implement an alternative platform for differential clock comparisons with a ``multiplexed'' one-dimensional (1D) OLC,
in which a movable 1D optical lattice is used to deterministically load up to 6 spatially-resolved ensembles of ultra-cold strontium atoms.
Fluctuating environmental perturbations such as black-body radiation (BBR),
magnetic and electric fields,
Doppler shifts from atomic motion or clock laser path length fluctuations,
and ac Stark shifts from the lattice and probe beams are common mode to first order between the ensembles,
significantly reducing atom-atom dephasing and uncertainty in the differential clock frequencies.
As a result, we observe atom-atom coherence times of 26 s with 2400 atoms per ensemble in a 1D lattice geometry, a factor of roughly 270 times longer than the measured atom-laser coherence time. This demonstrates that decoherence due to atomic collisions \cite{swallows_suppression_2011, martin_quantum_2013, campbell_fermi_degenerate_2017},
coupling of motion between the transverse and radial modes~\cite{blatt_rabi_2009},
and tunneling \cite{hutson_engineering_2019} need not limit 1D optical lattice clocks from achieving coherence well into the 10-second-scale. 
Furthermore, when combined with the high atom numbers, fast loading times, 
and low Raman scattering rates afforded by a weak vertical 1D lattice, 
these long atomic coherence times offer clock stabilities beyond those demonstrated in other platforms and geometries.

Through synchronous Ramsey interrogations with a single clock laser beam,
we measure a QPN-limited differential stability of $9.7(4)\times10^{-18}/\sqrt{\tau}$ between a pair of ensembles $0.6$~cm apart, and a fractional uncertainty of $8.9(3)\times 10^{-20}$ after 3.3 hours of averaging. These results illustrate that simultaneous differential clock comparisons enable record-setting stability and precision without requiring state-of-the-art mHz linewidth clock lasers, with important implications for applications that require portable or space-based clocks, such as relativistic geodesy and gravitational wave detection \cite{takano_geopotential_2016, kolkowitz_gravitational_2016}.
The same approach scales to the multiplexing of larger numbers of clock ensembles,
which we demonstrate by performing 15 unique pairwise clock comparisons between 6 atomic ensembles in parallel,
with relative stabilities below $3\times10^{-17}/\sqrt{\tau}$. 
Finally, we demonstrate the applicability of our approach to isotope shift comparisons by loading heterogeneous spatially resolved pairs of all four stable isotopes of strontium in the same lattice.

\begin{figure}[!t]
    \centering
    \includegraphics[width=0.95\textwidth]{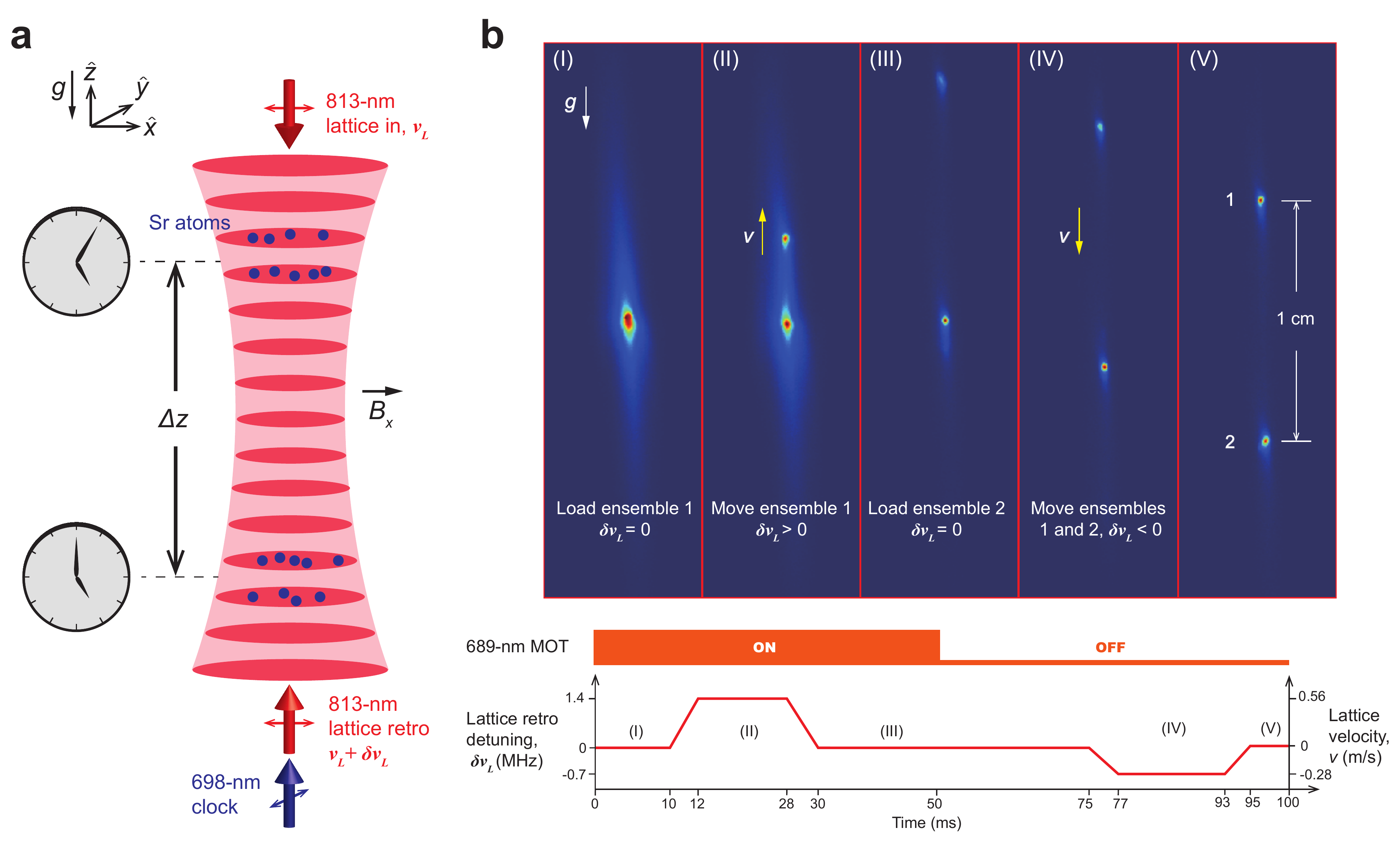}
    \caption{\textbf{Multiplexed optical lattice clock configuration and procedure for loading two ensembles.}
    \textbf{a}, schematic illustrating the multiplexed optical lattice clock concept.
    Two ensembles of strontium atoms separated by $\Delta h $ in height are prepared for simultaneous clock interrogation. 
    A bias magnetic field ($B_{x}$, about 2 G) along $\hat{x}$ defines the quantization axis.
    \textbf{b}, Top: images taken during loading stages (I-V). 
    (I): loading ensemble 1 into the lattice;
    (II) separating ensemble 1 from the original atomic cloud by accelerating the lattice along $+\hat{z}$;
    (III) loading ensemble 2 into the lattice;
    (IV) simultaneously moving both ensembles along -$\hat{z}$;
    (V) two stationary ensembles with a height difference of 1 cm are prepared for spin-polarization, in-lattice cooling, and clock interrogation.
    Bottom: timing sequence diagram including the 689-nm MOT and lattice retro detuning which governs the lattice loading and moving. }
    \label{fig1}
\end{figure}

The basic concept of the ``multiplexed'' OLC is illustrated in Fig.\ref{fig1}\textbf{a}.
A one-dimensional ``magic wavelength'' lattice ($\lambda_L = 813.4$~nm) is formed using an incoming beam (1.5 W power) focused to a 100 $\mu$m beam waist and a retro-reflected beam with a matching waist.
The lattice is orientated in $\hat{z}$ to suppress tunneling with the help of gravity,
which lifts the degeneracy between adjacent lattice sites~\cite{lemonde_optical_2005, hutson_engineering_2019}.
After passing through the science chamber, the beam is sent through 2 acousto-optic modulators (AOMs) that operate at opposite diffraction orders ($\pm$ 110 MHz),
and double-passed back with a ``cat's eye'' retro-reflector.
In this configuration,
the lattice retro frequency ($\nu_L + \delta \nu_L$) can be detuned from the incoming frequency ($\nu_L$) by varying the radio-frequency drive of the second AOM.
At zero detuning ($\delta \nu_L=0$), the lattice is a standing wave and the clock can be operated in the traditional fashion.
A constant detuning $\delta \nu_L$ results in a moving lattice with a velocity $v_L$
\begin{equation}
    v_L = \frac{1}{2}\lambda_L \cdot \delta \nu_L.
\end{equation}
If $\delta\nu_L$ is changed in time,
the lattice will accelerate at
\begin{equation}
    a_L = \frac{1}{2}\lambda_L \cdot (\partial{\delta \nu_L}/\partial{\delta t}),
\end{equation}
which in our apparatus can exceed $100~g$, 
mainly limited by the atomic temperature and lattice trap depth,
where $g~\approx$ 9.80 m/s$^2$ is the acceleration due to gravity.
The experimental procedure for loading two ensembles separated by 1 cm along $\hat{z}$ is shown in Fig.\ref{fig1}\textbf{b},
where 5 images are shown for loading and moving the lattice such that two ensembles with a tunable separation centered about the lattice beam waist can be prepared.
A resonant 461-nm imaging beam co-propagating along the lattice is used for fluorescence imaging  with an electron multiplied charged-coupled device (EMCCD).
In our apparatus a few thousand atom can be loaded in each ensemble with spatial separations ranging from $<1$ mm to $>1$ cm in under 100 ms.

%The lattice retro detuning $\delta\nu_L$ is initially set to zero to load atoms for the first ensemble from the 689~nm narrowline magneto-optical trap (MOT), which are then accelerated along the axis of the lattice by ramping $\delta\nu_L$ to a positive value. The lattice is then held at a constant velocity to move the atoms, followed by a ramp to $\delta\nu_L=0$. This allows us to separate the ensemble of atoms trapped in the lattice from the MOT, as shown in Fig.\ref{fig1}\textbf{b}-(II). A second ensemble is then loaded into the lattice from the MOT, at which point the MOT lasers and magnetic field gradients are switched off, and a second move is performed in the opposite direction from the first. Ultimately, this results in two ensembles with a tunable separation centered about the lattice beam waist (see Fig.\ref{fig1}\textbf{b}-(V)).

\begin{figure}[!ht]
    \centering
    \includegraphics[width=0.95\textwidth]{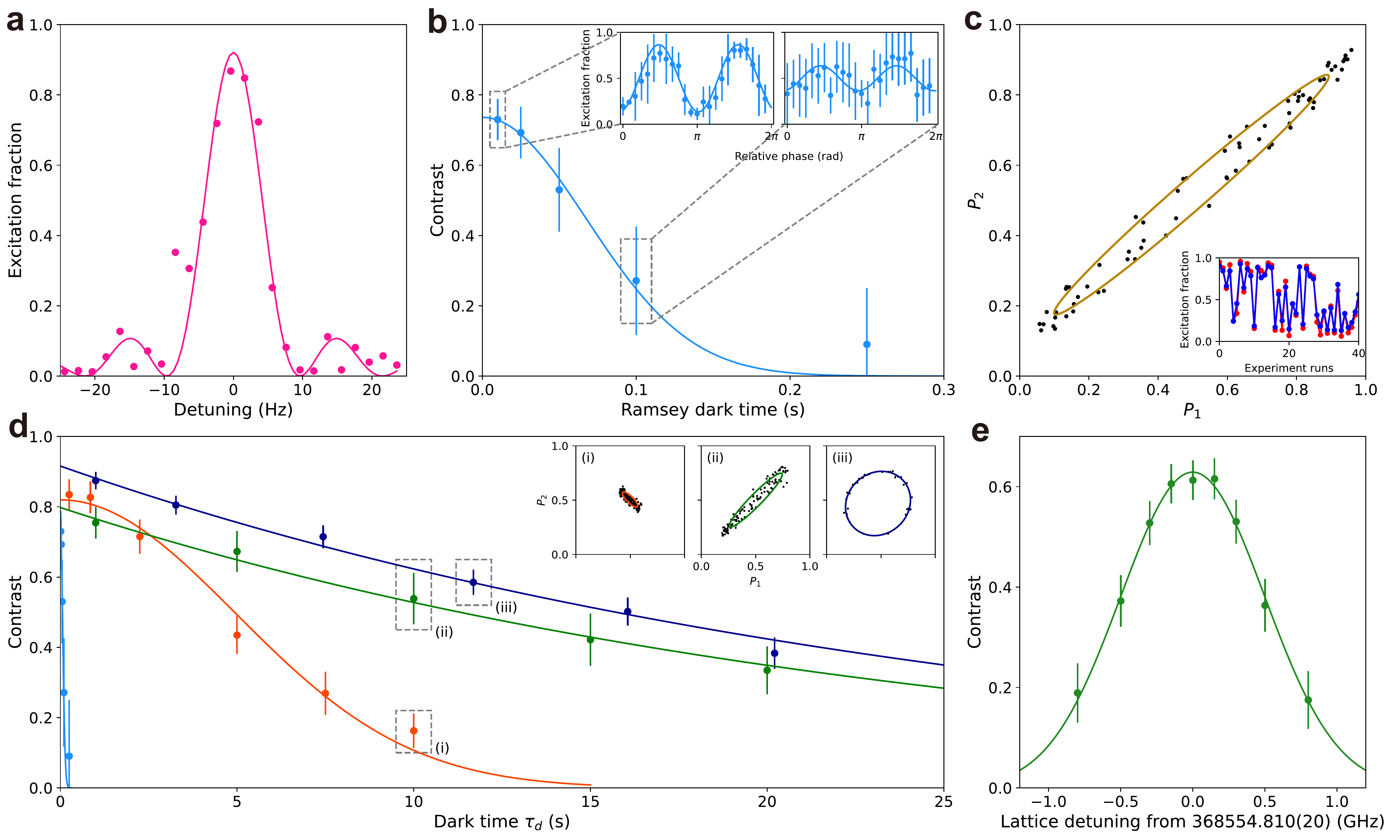}
    \caption{\textbf{Characterization of atom-atom coherence time by synchronous clock comparisons.}
    \textbf{a}, Rabi spectroscopy of a single ensemble with 90 ms $\pi$-pulse duration (pink circles), resulting in a fitted linewidth of 10.2(4) Hz (solid pink line).
    Each datum is taken with one experiment run without averaging.
    \textbf{b}, decay of Ramsey contrast taken with one ensemble.
    This is fitted to a Gaussian envelope which gives a $1/e$ coherence time of 96(24) ms. 
    Inset: Ramsey fringes at 10 and 100 ms dark times.
    \textbf{c}, parametric plot of excitation fractions in ensemble 1 ($P_1$) and ensemble 2 ($P_2$).
    Ellipse fitting is used to extract the differential phase (solid line).
    Inset: correlations in excitation fraction $P_1$ (red) and $P_2$ (blue) for the two regions.
    \textbf{d}, measurement of atom-atom coherence times in differential comparisons between two ensembles.
    Synchronous Ramsey interrogation on $\ket{^1S_0 , m_F=9/2}\leftrightarrow\ket{^3P_0, m_F=9/2}$ transition gives a $1/e$ coherence time of 6(1) s by fitting to a Gaussian envelope (red line), 
    while spin-echo on the same transition results in a coherence time of 24(5)~s with an exponential decay (green line).
    Synchronized Ramsey interrogation of the magnetically insensitive $\ket{^1S_0,m_F=5/2} \leftrightarrow \ket{^3P_0, m_F=3/2}$ clock transition results in a 26(2)~s atom-atom coherence time (dark blue line).
    Insets: representative parametric plots illustrating relative contrast.
    \textbf{e}, a new operational lattice magic wavelength for the $\ket{^1S_0 (m_F=5/2)}\leftrightarrow\ket{^3P_0(m_F=3/2)}$ transition is found by measuring the contrast of synchronous Ramsey interrogations at 20~$E_{\rm rec}$ trap depth and 10 s dark time as a function of lattice frequency.
    }
    \label{fig2}
\end{figure}

Before interrogating the clock transition,
we spin-polarize the samples into ${}^1S_0~(F=9/2$, $m_F=9/2)$ states,
where $m_F$ is the projection of the total angular momentum $F$ along the quantization axis defined by the applied bias magnetic field $B_x$ ($\sim$ 2 G),
and perform in-lattice-cooling to remove heating after lattice acceleration.
We then interrogate the ${}^1S_0\leftrightarrow{}^3P_0$ clock transition with a 698~nm clock laser that is referenced to a 12 cm ultra-low-expansion (ULE) cavity.
Limited by the ULE cavity,
we expect a local oscillator linewidth of $\sim1$~Hz and a linear drift rate $\sim1$ Hz/s before drift cancellation.
This is orders of magnitude worse than  state-of-the-art cavities such as cryogenic ultra-stable silicon cavities~\cite{matei_15text_2017, zhang_ultrastable_2017, robinson_crystalline_2019},
with measured linewidths of $8$~mHz and linear drift rates $<1$~mHz/s~\cite{milner_demonstration_2019},
and has been used to demonstrate differential stabilities at low $10^{-19}$ level~\cite{campbell_fermi_degenerate_2017, marti_imaging_2018, young_half_minute_scale_2020}.

To characterize the limitations placed on the coherent interrogation time by the clock laser linewidth, we first study each ensemble independently.
A representative Rabi spectrum with a 10 Hz linewidth is shown in Fig.\ref{fig2}\textbf{a},
where a $\pi$-pulse of 90 ms duration is used to drive the ${}^1S_0(m_F=9/2)\leftrightarrow{}^3P_0(m_F=9/2)$ (denoted as $\ket{g, 9/2}\leftrightarrow\ket{e,9/2}$ below) transition.
Further increasing the pulse duration results in a reduction of excitation fraction but does not reduce the linewidth due to laser frequency noise.
The atom-laser coherence is also measured via Ramsey spectroscopy (Fig.\ref{fig2}\textbf{b}) by varying the relative phase between the two $\pi/2$-pulses.
While the fringe contrast decays with a Gaussian time constant of 96(24) ms,
the variance of the Ramsey signal remains high at 100 ms (Fig.\ref{fig2}\textbf{b}, inset2),
implying that the atom-atom coherence time can be longer.
The loss of atom-laser coherence is due to the finite coherence time of the clock laser, and manifests itself as a randomized phase of the second $\pi/2$-pulse. 
However, when the two atom ensembles are probed simultaneously, the relative atomic phase is preserved and reflected in a high degree of correlation between the excitation fractions of the two ensembles (Fig.\ref{fig2}\textbf{c}, inset).
This can be further clarified in a parametric plot of excitation fractions for ensembles 1 and 2,
which fall on an ellipse with an opening angle determined by the differential Ramsey phase acquired between the two ensembles (Fig.\ref{fig2}\textbf{c}). 
This differential phase is a measure of the detuning between the two atomic ensembles,
and contains information about all of the differential frequency shifts experienced by the spatially separated ensembles, 
including differential linear and quadratic Zeeman shifts due to magnetic field gradients, 
differential dc Stark shifts due to electric field gradients,
differential ac Stark shifts from the lattice and probe light due to differing field intensities at the two ensembles,
differential BBR shifts due to temperature gradients across the apparatus,
and the gravitational redshift due to general relativity.
In our apparatus we find that the dominant shifts are the linear and quadratic Zeeman shifts due to residual magnetic field gradient of $B_x$ along $\hat{z}$ with amplitude $\sim$ 15 mG/cm.
At $1$~cm this corresponds to a detuning between the $\ket{g, 9/2}\leftrightarrow\ket{e, 9/2}$ clock transitions of the two ensembles of 7.5 Hz due to the differential linear Zeeman shift,
and a differential quadratic Zeeman shift of 14 mHz at a bias field of $B_x=2$~G.

To investigate the atom-atom coherence times, we perform synchronized Ramsey interrogation between two ensembles.
As pointed out in prior work~\cite{campbell_fermi_degenerate_2017, hutson_engineering_2019, norcia_seconds_scale_2019, young_half_minute_scale_2020},
a shallower lattice trap depth is desired for second-scale coherent interrogation to minimize Raman scattering out of the $\ket{^3P_0}$ state~\cite{dorscher_lattice_induced_2018}.
We operate at a lattice depth of 20~$E_{\rm rec}$ with a measured excited state lifetime of 13(2) s.
However, 
when probing the $\ket{g, 9/2}\leftrightarrow\ket{e, 9/2}$ transition we find that the contrast decays with a Gaussian time constant of 6(1)s,
which is below the excited state lifetime and suggests inhomogeneous broadening during Ramsey free evolution.
This is consistent with the expected broadening due to the magnetic field gradient across $\hat{z}$ of 200 $\mu$m finite spatial extent of each region,
corresponding to a detuning of 150(10) mHz from the top to the bottom of each ensemble.

To confirm that the magnetic field gradient is limiting the atom-atom coherence time,
we perform a ``spin-echo'' measurement,
where a $\pi$ pulse is applied after half the dark time $\tau_d$ to cancel out the differential phase accumulation between the two ensembles for constant detunings.
With ``spin-echo'',
we observe an exponential time constant 24(5)~s.
To take full advantage of this longer available coherence time,
we therefore switch to interrogating the $\ket{g,5/2}\leftrightarrow\ket{e, 3/2}$ transition,
with a magnetic field sensitivity of about 22.4 Hz/G,
22 times smaller than $\ket{g, 9/2}\leftrightarrow\ket{e, 9/2}$ transition \cite{boyd_nuclear_2007,oelker_demonstration_2019}.
This is done by coherently transferring spin-polarized atoms from $\ket{g, 9/2}$ to $\ket{e, 3/2}$ state via three sequential $\pi$-pulses on resonance with $\ket{g, 9/2} \leftrightarrow \ket{e, 7/2}$, $\ket{e, 7/2}\leftrightarrow \ket{g, 5/2}$ and $\ket{g, 5/2}\leftrightarrow \ket{e, 3/2}$ transitions, respectively (see Extended Data Fig.\ref{figED.energy_levels}).
About 70\% of atoms are transferred to the $\ket{e, 3/2}$ state, 
which is mainly limited by the $\pi$-pulse fidelity.
For this transition the magnetic field gradient across each ensemble can be expected to contribute a detuning of only 7(1) mHz, and therefore no longer contributes dephasing on timescales limited by the Raman scattering.

Due to the tensor ac Stark shift, the $\ket{g, 5/2}\leftrightarrow \ket{e, 3/2}$ transition will have a different lattice operational magic wavelength (where the scalar and tensor ac Stark shifts sum to zero) than the $\ket{g, 9/2}\leftrightarrow\ket{e, 9/2}$ transition.
By scanning the lattice laser frequency over a range of $\pm 800$ MHz,
we find a lattice frequency that maximizes the contrast for $\ket{g, 5/2}\leftrightarrow\ket{e,3/2}$ transition at 368,554,810(30) MHz (see Fig.\ref{fig2}\textbf{e}),
where the uncertainty comes from both the error in the fitting and accuracy of the wave-meter (10 MHz).
We observe an $1/e$ atom-atom coherence time of $26(2)$~s on $\ket{g, 5/2}\leftrightarrow\ket{e,3/2}$ transition,
which is consistent with the ``spin-echo'' measurement on $\ket{g, 9/2}\leftrightarrow\ket{e,9/2}$ transition,
and is about twice the measured excited state lifetime (see Supplementary Information),
implying that we are primarily limited by Raman scattering. We note that here the atom-atom coherence time refers only to the lifetime of the synchronized Ramsey contrast for atoms remaining in the lattice at the end of the sequence, and therefore does not include atom loss due to heating and background gas collisions.

\begin{figure}[!t]
    \centering
    \includegraphics[width=0.95\textwidth]{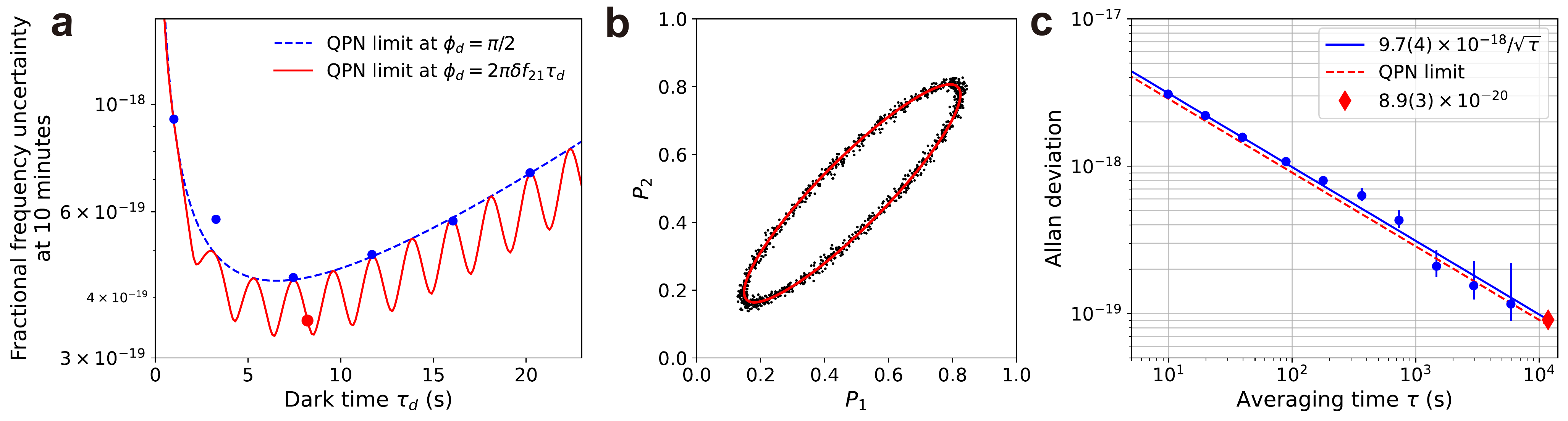}
    \caption{{\textbf{High differential stability with multiplexed Ramsey interrogation.}}
     \textbf{a}, measured fractional frequency uncertainties after 10 minutes averaging time (blue points) by choosing dark times such that the differential phase $\phi_d$ is close to an odd multiple of $\pi/2$.
     Blue dashed line is the expected QPN limit at a fixed offset phase $\pi/2$.
     Red solid line is QPN accounting for $\phi_d = 2\pi\delta f_{\textrm{21}}\tau_d$,
     where $\delta f_{\textrm{21}}$ is the frequency detuning between 2 ensembles and $\tau_d$ is the Ramsey dark time.
     The oscillation feature comes from the fact that QPN is maximized at $\pi/2$ and minimized at 0 or $\pi$ rad when contrast is below 1.
     \textbf{b}, parametric plots of $P_2$ versus $P_1$ (black points) taken at $\tau_d$ = 8.205 s (red point in \textbf{a}) with 1193 experimental runs for a total measurement time of 11800 s.
     A least-squares method (see Supplementary Information) is used to fit an ellipse to the data (red solid line).
     \textbf{c}, the corresponding Allan deviation (blue points) from the data shown in \textbf{b} extracted via jackknifing~\cite{marti_imaging_2018,young_half_minute_scale_2020}.
     The differential clock comparison averages down with a stability of $9.7(4)\times10^{-18}/\sqrt{\tau}$ (blue solid line), which matches the QPN limit (red dash line) for the independently measured number of atoms in each ensemble ($N_a\approx2400$),
     and reaches a differential fractional frequency uncertainty of $8.9(3)\times10^{-20}$ after 3.3 hours of averaging (red diamond).
     }
    \label{fig3}
\end{figure}

To characterize the stability of the multiplexed OLC,
we perform a synchronous clock comparison between 2 ensembles separated by 0.6 cm on $\ket{g, 5/2}\leftrightarrow\ket{e, 3/2}$ transition.
Due to competition between a $1/e$ decay of contrast and $1/\sqrt{\tau_d}$ scaling of QPN,
the optimal Ramsey dark time can be found by comparing fractional frequency uncertainties at different dark times (Fig.\ref{fig3}\textbf{a}, blue points),
which are chosen such that the differential phase $\phi_d$ is close to an odd multiple of $\pi/2$ to minimize biased error from ellipse fitting~\cite{marti_imaging_2018,young_half_minute_scale_2020}.
The measurement agrees with the QPN limit at a fixed offset phase of $\pi/2$ (Fig.\ref{fig3}\textbf{a}, blue dashed line),
which suggests an optimal $\tau_d$ at 7.5 s.
However,
due to phase evolution of $\phi_d= 2\pi\delta f_{\textrm{21}}\tau_d$ at a comparable time scale to the dark times,
where $\delta f_{\textrm{21}}$ is the frequency difference between two ensembles,
an additional differential-phase dependent scale factor must be included in the expected QPN limit when contrast is below $1$ (see Supplementary Information).
As a result,
QPN is maximized when the offset phase is at $\pi/2$,
and minimized at $0$ or $\pi$.
This implies one can benefit in sensitivity by trading off for biased ellipse fitting (Fig.\ref{fig3}\textbf{a}, red line).
Therefore,
we choose $\tau_d=8.205$ s such that the offset phase is $\sim$ 0.44 rad,
at which the biased error is bound to below 3\% and can be easily compensated for (see Supplementary Information).
Fig.\ref{fig3}\textbf{b} shows the measurement taken with 1193 experiment runs recorded in 3.3 hours,
and the corresponding fitted ellipse.
The fit yields a net phase shift of about 12.130(2) rad,
or a frequency difference of 235.29(4) mHz between the 2 ensembles.
The overlapping Allan deviation is computed and plotted in Fig.\ref{fig3}\textbf{c},
with a relative stability of $9.7(4)\times 10^{-18}/\sqrt{\tau}$ in agreement with the QPN limit (red dashed line),
and a fractional frequency uncertainty of $8.9(3)\times 10^{-20}$ at the full 3.3 hours of averaging time.
This demonstration of precision below the $10^{-19}$ level with a rack-mounted, commercially-available local oscillator with a stability of $1\times10^{-15}$ at 1 s is encouraging for future applications that require portable or spaced-based clocks such as relativistic geodesy, and  gravitational wave detection~\cite{takano_geopotential_2016, grotti_geodesy_2018, takamoto_test_2020, kolkowitz_gravitational_2016, burt_demonstration_2021}.

\begin{figure}[!t]
    \centering
    \includegraphics[width=0.95\textwidth]{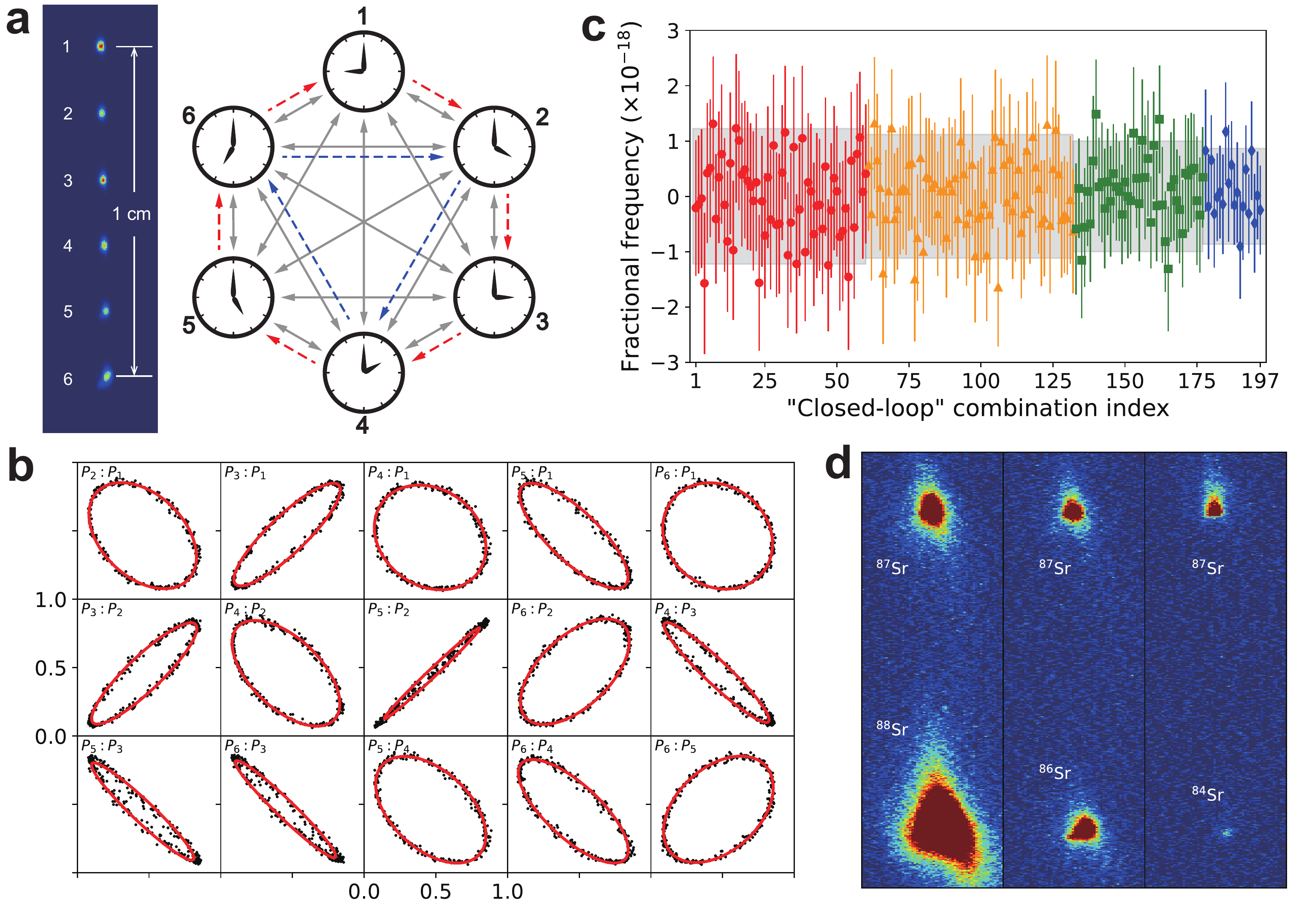}
    \caption{\textbf{Prospects for multiplexed OLC comparisons.}
    \textbf{a}, CCD image of 6 equally-spaced ensembles loaded and interrogated simultaneously,
    which corresponds to 15 unique pairwise differential clock comparisons (grey double arrows with solid lines).
    The red arrows with dashed lines form a representative ``closed-loop'' (1, 2, 3, 4, 5, 6) for clock comparisons with 6 ensembles. 
    The sum frequency within this loop is given by $\delta f_{21} + \delta f_{32} + \delta f_{43} +\delta f_{54} + \delta f_{65} -\delta f_{61}$,
    where $\delta f_{ij}= f_j - f_i$ is the differential frequency between ensemble $i$ and $j$, and should sum to zero within the uncertainty of the measurement.
    Similarly,
    the blue arrows with dashed lines form a ``closed-loop'' (2, 4, 6) for clock comparisons with 3 ensembles.
    \textbf{b}, all 15 parametric plots for simultaneous pairwise comparisons with 6 ensembles,
    each comparison averages down to a fractional frequency uncertainty of about $5\times10^{-19}$ in an hour.
    \textbf{c}, ``Closed-loop'' analysis of 15 pairwise comparisons as a self-consistency check.
    Each datum corresponds to the summed fractional frequency of the pairwise comparisons within the loop.
    This includes a total of 197 unique loops with 6 ensembles (red, 60 combinations),
    5 ensembles (orange, 72 combinations),
    4 ensembles (green, 45 combinations),
    and 3 ensembles (blue, 20 combinations).
    The shaded grey areas represent uncertainty windows of $5\times10^{-19}$ scaled by $\sqrt{6}$, $\sqrt{5}$, $\sqrt{4}$, and $\sqrt{3}$, respectively.
    \textbf{d}, demonstration of simultaneous loading of spatially-resolved heterogeneous pairs of isotopes in a single experiment run (three separate representative experiments are shown covering all four stable isotopes of Sr). 
    Each isotope in a pair is imaged individually by shifting the 461-nm probe beam onto resonance sequentially.
    The color-map is kept on the same scale for all three images, with the atom numbers for the bosonic isotopes consistent with their relative isotopic abundance (${}^{88}$Sr - 82.6\%, ${}^{87}$Sr - 6.9\%\, ${}^{86}$Sr - 9.9\%, ${}^{84}$Sr - 0.6\%).
    }
    \label{fig4}
\end{figure}

We demonstrate the scability of the multiplexed OLC technique by moving from pairs to larger numbers of ensembles. 
This is achieved by modifying the loading sequence shown in Fig.\ref{fig1}\textbf{b} and repeating the lattice acceleration - loading cycle several times.
A representative CCD image is shown in Fig.\ref{fig4}\textbf{a},
where 6 ensembles are equally distributed with 0.2 cm spacing.
Each ensemble has about 500 atoms and the total lattice loading time is less than 100 ms.
Simultaneous clock interrogation and read-out results in 15 unique pairwise clock comparisons from a miniature network consisting of 6 ``clocks'' (Fig.\ref{fig4}\textbf{b}).
Each comparison averages down with a slope below $3\times10^{-17}/\sqrt{\tau}$,
and reaches a fractional frequency uncertainty of roughly $5\times10^{-19}$ after 1 hour of averaging.
To verify that the 15 pairwise comparisons are self-consistent,
we perform a ``closed-loop'' analysis where each loop contains 3 or more ``clocks''.
This results in a total number of 197 possible unique combinations after removal of cyclic degeneracy (see Supplementary Information for details).
While the result of an individual pairwise comparison contains both differential frequency and measurement noises,
the sum of the frequency differences between pairs around a ``closed-loop'' should always be zero, leaving only measurement noise
regardless of any spatial gradients or systematic shifts.
The ``closed-loop'' analysis shows good agreement with zero within the expected uncertainty of $5\times10^{-19}$ scaled by $\sqrt{6}$, $\sqrt{5}$, $\sqrt{4}$, and $\sqrt{3}$ for ``closed-loop'' combinations composed of 6, 5, 4, and 3 ensembles, respectively (see Fig.\ref{fig4}\textbf{c}).

The self-consistency within the 15 pairwise comparisons confirms the validity and accuracy of extracting differential clock detunings by ellipse fitting to synchronized Ramsey measurements, and represents a critical step towards mapping out spatial gradients across the lattice.
The measured detunings between ensemble pairs contain information about the spatial profiles of the magnetic field gradient, the lattice beam, thermal gradients, and electric field gradients, as well as residual differential shifts due to differences in atom density and temperature between the ensembles.
As an example,
we evaluate the differential density shifts between ensemble pairs by varying the relative atom numbers,
and thus the differential density (see Extended Data Fig.\ref{figED.DensityShifts}).
At a typical lattice trap depth $U$ of $20~E_{\textrm{rec}}$ and a conservative 100(25) atom number difference,
the differential density shift is $-8(2)\times10^{-19}$.
By varying the lattice trap depths,
we further observe the expected density shift scaling with depth of $U^{5/4}$ as reported in \cite{swallows_operating_2012, bothwell_jila_2019}.
By working at shallower lattices and actively controlling atom loading,
differential density shifts with uncertainties at the $10^{-20}$ level should be feasible.
This example highlights the utility of the multiplexed technique for mapping out and evaluating systematic effects~\cite{Beloy_Faraday_2018,marti_imaging_2018}.
A thorough evaluation of all of the contributing differential systematic shifts in our apparatus is currently underway.

Finally,
precision isotope shift measurements have recently been proposed as a novel method to search for new physics beyond the Standard Model~\cite{king_w_h_isotope_1984, delaunay_probing_2017, berengut_probing_2018, flambaum_isotope_2018, miyake_isotope_shift_2019, counts_evidence_2020}.
Neutral strontium, with four stable isotopes ($^{88}$Sr, $^{87}$Sr, $^{86}$Sr and $^{84}$Sr) and narrow-line clock transitions, is a good candidate when combined with measurements of the clock transitions in the Sr$^+$ ion ~\cite{flambaum_isotope_2018}.
We demonstrate the capability to sequentially load different spatially-resolved strontium isotopes into the lattice with pairs comprising all four stable isotopes of strontium (see Fig.\ref{fig4}\textbf{d}).
We avoid scattering and heating during loading of the second isotope by shelving the first trapped isotope, ${}^{87}$Sr,
in the $^3P_0$ state.
Due to the lack of hyperfine states for bosonic isotopes,
a greater magnetic field ($\sim$ 20 G) is required to mix $^3P_1$ into $^3P_0$ state~\cite{taichenachev_magnetic_2006,madjarov_atomic_array_2019, norcia_seconds_scale_2019,young_half_minute_scale_2020},
and allow the doubly forbidden ${}^1S_0\leftrightarrow{}^3P_0$ clock transition,
and is left for future work.

In conclusion, in this work
we implement an alternative platform for differential clock comparisons using spatially resolved atom ensembles trapped in a single 1D optical lattice.
We demonstrate long atomic coherence times (26 s) with large atom numbers (2400 atoms per ensemble) in a shallow ($20~E_{\textrm{rec}}$) vertical optical lattice using a 1 Hz linewidth clock laser.
In a comparison between two regions we achieve a QPN-limited differential stability of $9.7(4)\times10^{-18}/\sqrt{\tau}$,
and a fractional frequency precision of $8.9(3)\times10^{-20}$ after 3.3 hours of averaging. We take advantage of the multiplexed nature of our apparatus to demonstrate a miniaturized clock network consisting of 6 atom ensembles,
resulting in 15 unique pairwise clock comparisons with relative stabilities below $3\times10^{-17}/\sqrt{\tau}$. Finally, we demonstrate the capability to simultaneously load heterogeneous pairs of all four stable isotopes of strontium into spatially-resolved ensembles in the lattice.
Common-mode rejection of dephasing from environmental fluctuations and local oscillator noise make the multiplexed optical lattice clock platform well-suited for exploring the use of spin-squeezing~\cite{kitagawa_squeezed_1993, bouchoule_spin_2002, gil_spin_2014, van_damme_impacts_2021, pedrozo_2020_entanglement} to push the differential stability below the QPN limit.
Full characterization of systematic effects such as differential BBR, Stark, and  Zeeman shifts will open up possibilities for studying relativistic geodesy at the sub-cm scale and other novel tests of general relativity.
Similarly,
we anticipate that extensions of this technique to other existing 1D-OLCs will be straightforward, enabling high-stability characterization of limiting clock systematics such as magnetic, electric, and thermal gradients.

\vskip 0.2in
\textit{Authors' note:}
While performing the work described here, 
we became aware of complementary work in which record stability and precision clock comparisons were performed between sub-regions within a single atomic ensemble in a vertical 1D lattice using an ultra-narrow linewidth local oscillator stabilized to a cryogenic single crystal silicon cavity~\cite{bothwell_jila_2021}.

\printbibliography[keyword={maintext},title={${}$}]

\pagebreak

\section*{\hfil Methods \hfil}

\addto\captionsenglish{\renewcommand{\figurename}{Extended Data Fig.}}

\subsection*{\hfil Initial loading and trapping \hfil}

The experiment starts by capturing atoms from a thermal atomic beam in a 3D-MOT operating on the ${}^1S_0\leftrightarrow{}^1P_1$ transition at 461-nm,
which has a linewidth of 32 MHz.
The atom number in the 461-nm MOT is typically $2\times10^6$ for ${}^{87}$Sr,
with a temperature of $\sim$1 mK.
The sample is further cooled by transferring from the 461-nm MOT into a 689-nm MOT via the 7.5 kHz wide ${}^1S_0\leftrightarrow{}^3P_1$ transition.
After broad-band (BB) and single-frequency (SF) 689-nm MOT stages,
about $2\times10^5$ atoms are left with a temperature of $\sim2~\mu$K.
The optical lattice is kept on during the entire experiment,
and about $1\times10^5$ spin-mixed atoms are transferred into the optical lattice by switching off the 689-nm MOT.
The optical lattice light is generated by a Ti:Sapphire laser (MSquared Soltis),
diffracted by an AOM operating at 80 MHz,
and the negative first diffraction order is delivered to the experiment table through a polarization-maintaining (PM) fiber.
The lattice laser intensity is servoed on the AOM by picking off the lattice beam after the science chamber.
The lattice laser frequency is digitally locked to a wave-meter (High-Finesse, WS-70) that is calibrated using the ${}^1S_0\leftrightarrow{}^3P_0$ clock transition of ${}^{87}$Sr,
which is known to an accuracy better than 1 Hz~\cite{GKCampbell_absolute_2008}.

\subsection*{\hfil Accelerating lattice \hfil}
In order to generate an accelerating lattice, a tunable frequency difference between the incoming and retro-reflected lattice beams while maintaining their spatial overlap is required.
To realize this,
the incoming lattice beam is re-shaped with a pair of telescope lenses after the science chamber,
and is subsequently sent through two AOMs (lattice AOM 1 and lattice AOM 2) operating at $\mp110$ MHz (see Extended Data Fig.\ref{figED.clock_lattice_path}).
A ``cat's eye'' retro-reflector consisting of an $100$~mm lens and a high-reflection mirror is used to retro-reflect the lattice beam and double-pass the AOMs.
The power of the retro-reflection beam is about 50\% compared to the incoming beam,
mainly limited by the AOM diffraction efficiencies ($\sim$~90\% per single pass),
and optical losses in the path.
Two direct-digital-synthesizers (DDS's, Moglabs XRF421) synchronized in phase are used to drive the lattice AOMs.
The DDS-2 that drives the lattice AOM 2 is programmed to perform acceleration after receiving an external trigger signal,
which is typically 5 ms after the SF 689-nm MOT stage has started.
The DDS-2 frequency is then stepped over 4000 values for a 2 ms ramp with
 an update rate of 500 ns.
About 80\% of the total atoms survive after lattice acceleration and deceleration.
The ``cat's eye'' configuration is critical for ramping the lattice frequency while preserving the lattice overlapping,
which is monitored through the rejection port of an optical isolator before the fiber.
We observe negilible power loss when detuning the retro-lattice frequency by as much as $\pm$10 MHz,
which is more than sufficient to prepare ensembles separated by 1 cm in the experiment.

\subsection*{\hfil State preparation, cooling and read-out \hfil}

For ${}^{87}$Sr,
a 689~nm laser beam propagating perpendicular to the lattice is applied to spin-polarize atoms into $\ket{m_F=\pm 9/2}$ hyperfine state manifold via the ${}^1S_0(F=9/2)\leftrightarrow{}^3P_1(F=9/2)$ transition.
We then perform sideband cooling on the 689~nm ${}^1S_0\leftrightarrow{}^3P_1$ transition to remove phonons after lattice acceleration,
and adiabatically ramp down the lattice trap depth from 60 $E_{\rm rec}$ down to below 20 $E_{\rm rec}$,
where $E_{\rm rec}\approx 3.5 ~h\cdot$kHz is the lattice photon recoil energy.
To prepare ensembles into $m_F = \pm3/2$ states,
we coherently transfer the populations via three $\pi$-pulses on resonant with the $\ket{{}^1S_0,m_F= \pm9/2}\leftrightarrow\ket{{}^3P_0, m_F=\pm 7/2}$,
$\ket{{}^3P_0,m_F= \pm7/2}\leftrightarrow\ket{{}^1S_0, m_F=\pm 5/2}$, 
and $\ket{{}^1S_0,m_F= \pm5/2}\leftrightarrow\ket{{}^3P_0, m_F=\pm 3/2}$ transitions.

To detect the excitation fraction of each ensemble in parallel,
we first read out $^1S_0$ ground state ($\ket{g}$) populations with a 1-ms imaging pulse with a co-propagating 461-nm laser beam along the lattice,
and the fluorescence is collected on the EMCCD.
The probe beam also clears out the population in $\ket{g}$.
The remaining populations in $^3P_0$ excited state ($\ket{e}$) are simultaneously transferred back to $\ket{g}$ via repump pulses on the ${}^3P_0\leftrightarrow{}^3S_1$ and ${}^3P_2\leftrightarrow{}^3S_1$ transitions,
and imaged with a second imaging pulse.
A reference image is taken with a final imaging pulse without any atoms for background subtraction.
Excitation fractions of each ensemble can be extracted by post-selecting regions-of-interest within the images,
and normalized excitation fraction is given by $P_{e,n} = (N_{e,n}-N_{bg, n})/(N_{e,n}+N_{g,n}-2N_{bg, n})$,
where $n$ is the $n$-th ensemble,
$N_{g/e}$ is the atom number for $\ket{g/e}$ after calibration,
and $N_{bg, n}$ is the background.

\subsection*{\hfil Clock laser beam path \hfil}

The rack-mount clock laser (Menlo Systems, Optical Reference System) is referenced via Pound-Drever-Hall locking to a 12 cm ULE cavity,
which is temperature controlled at the zero-crossing setpoint 15.77 $^{\circ}$C.
An double-passed AOM before fiber coupling into the ULE cavity is used for linear drift cancellation.
A typical linear drift rate from 0.2 to 1 Hz/s is observed,
and a residual drift of less than 0.01 Hz/s can be achieved upon calibration based the clock transition resonance.
The clock laser beam is delivered to the experiment table through a 5 m PM fiber,
with an output power of $\sim2$ mW,
and is subsequently sent to the clock AOM operating at +110 MHz to steer the laser frequency to be on resonant with the ${}^1S_0\leftrightarrow{}^3P_0$ clock transition.
The clock beam is focused down to a beam waist of about $500~\mu$m centered at the lattice,
which is about 5 times the lattice beam waist to both ensure homogeneity for atoms populated radially and multiple ensembles distributed axially along the lattice.

To cancel fiber phase noise and residual Doppler noise induced by vibrations of the fiber and the optical lattice,
the zeroth diffraction order of the clock AOM is referenced on the lattice retro-reflection mirror for Doppler cancellation.
Depending on the clock transition ($\ket{^1S_0, 9/2} \leftrightarrow \ket{^3P_0, 9/2}, \pi$-transition or $\ket{^1S_0, 5/2} \leftrightarrow \ket{^3P_0, 3/2}, \sigma$-transition),
the first diffraction order is overlapped with the lattice beam by either using a long-pass dichroic beamsplitter (for $\pi$ transition),
or using the reflection port of a polarized beamsplitter (for $\sigma$ transition, see Extended Data Fig.\ref{figED.clock_lattice_path}).
While this configuration leaves an uncompensated lattice path of about 75 cm,
we observe no significant impacts to the stability as inferred from the synchronous Ramsey interrogations.

\subsection*{\hfil Clock interrogations \hfil}

After loading atomic ensembles in the lattice and optical pumping into $m_F = \pm 9/2$ stretched states,
the clock transition is interrogated under a bias magnetic field of approximately 2~G.
The first diffraction order of the clock AOM is used to address the clock resonance.
To circumvent thermal effects in the AOM crystal,
the clock pulses are generated via jumping the AOM frequency from 10 MHz off-resonant to on-resonant,
instead of switching on and off the AOM.
The differential Bragg diffraction angle at 10 MHz frequency difference introduces a deflection of more than 0.5 cm at the ensembles,
in addition with an optical shutter that blocks any residual clock light,
which ensures the clock beam is kept off of the atoms during Ramsey free evolution.

The Rabi spectroscopy shown in Fig.\ref{fig2}a is taken on the $\ket{{}^1S_0, m_F=9/2}\leftrightarrow \ket{{}^3P_0, m_F= 9/2}$ transition at a lattice trap depth of 20 $E_{\textrm{rec}}$,
and with a $\pi$-pulse duration of about 90 ms,
which corresponds to $\approx 2\pi\times 5.6$ Hz Rabi frequency.
A neutral-density filter (3.5 optical density) is used to attenuate the clock laser power and ensure the resulting Rabi linewidth remains Fourier limited.
The data in Fig.\ref{fig2}\textbf{a} is taken within a total measurement time of less than a minute without averaging.

For Ramsey spectroscopy and ``spin-echo'' on the $\ket{{}^1S_0, 9/2}\leftrightarrow \ket{{}^3P_0, 9/2}$ transition,
the $\pi/2$-pulse duration is about 0.75 ms ($2\pi \times 333$ Hz in Rabi frequency).
For Ramsey spectroscopy on $\ket{{}^1S_0, 5/2}\leftrightarrow \ket{{}^3P_0, 3/2}$ transition,
the atoms must be transferred from the initial optically-pumped $m_F=\pm 9/2$ state to the $m_F=\pm3/2$ state.
This is achieved by using three consecutive $\pi$-pulse of about 4.5, 3.5, and 3.0 ms ($2\pi \times 111$ Hz, $2\pi \times 142$ Hz and $2\pi \times 167$ Hz in Rabi frequencies) to address the $\ket{{}^1S_0, 9/2}\leftrightarrow \ket{{}^3P_0, 7/2}$,
$\ket{{}^3P_0, 7/2}\leftrightarrow \ket{{}^1S_0, 5/2}$,
and $\ket{{}^1S_0, 5/2}\leftrightarrow \ket{{}^3P_0, 3/2}$ transitions,
respectively~(see Extended Data Fig.\ref{figED.energy_levels}).
The difference in pulse durations is a result of the different matrix elements for the 3 transitions.
Each $\pi$-pulse is followed by a ``clean-up'' pulse on resonant with 461-nm ${}^1S_0\leftrightarrow{}^1P_1$ transtition (679-nm ${}^3P_0\leftrightarrow{}^3S_1$ and 707-nm ${}^3P_2\leftrightarrow{}^3S_1$ repump transitions) to clean remaining populations on the ground (excited) state due to imperfect spin-polarization and $\pi$-pulses.
The nearby clock resonances from the final $m_F=3/2$ state,
for example,
the $\ket{^3P_0, m_F=3/2} \leftrightarrow \ket{^1S_0, m_F=3/2} $, $\pi$-transition and 
the $\ket{^3P_0, m_F=3/2} \leftrightarrow \ket{^1S_0, m_F=1/2} $, $\sigma$-transition,
can be eliminated by both applying a large bias magnetic field of 2~G to induce larger separation between $\sigma^+$ and $\sigma^-$ transitions,
and fine-alignment of the bias field orientation to suppress the unwanted $\pi$-transition.
After preparing atoms on $m_F=3/2$ state,
Ramsey spectroscopy is taken with $\pi/2$-pulses of 1.5 ms duration and dark times of up to 20 s.

\subsection*{\hfil Experimental procedure \hfil}
The procedure and timing diagram for loading, lattice acceleration, cooling, clock interrogation, and imaging is shown in Extended Data Fig.\ref{figED.timing_diagram}.
It takes 400 ms to load thermal atoms into the 461-nm MOT,
450 ms to cooling in the BB 689-nm MOT,
and 50 ms to further cool down to $\sim2~\mu$K by holding in the SF 689-nm MOT.
In the SF MOT stage,
the lattice is accelerated by linearly ramping the lattice AOM 2 frequency to load multiple ensembles within less than 100 ms.
This is then followed by spin-polarization, in-lattice cooling,
and adiabatic ramping down of the lattice trap depth within less than 200 ms.
An extra 100 ms is spent on coherent transfer from $m_F=9/2$ to $m_F=3/2$ hyperfine state when interrogating the $\ket{^1S_0, 5/2}\leftrightarrow\ket{^3P_0, 3/2}$ transition.
The imaging sub-sequence usually takes 150 ms and consists of 3 steps.
A first imaging pulse on resonant with the $\ket{{}^1S_0}\leftrightarrow\ket{{}^1P_1}$ transition at 461-nm is used to measure the population of atoms in the ground state and heated out of the trap.
This is then followed by a repumping pulse on resonant with the $\ket{{}^3P_0}\leftrightarrow\ket{{}^3S_1}$ and $\ket{{}^3P_2}\leftrightarrow\ket{{}^3S_1}$ transitions at 679-nm and 707-nm, respectively,
to transfer the excited state populations into the ground state,
which is subsequently read-out via a second imaging pulse.
A third imaging pulse is employed to measure the background.
The above sample trapping, cooling, state-preparation and read-out times contribute to a typical dead time of 1.6~s per experimental cycle.
This yields an 84\% duty-cycle for an 8.2 s Ramsey interrogation.

\subsection*{\hfil Ellipse fitting bias correction \hfil}

To determine the differential phase $\phi_d$ between ensemble pairs accumulated during the clock interrogation,
a least square method is applied for ellipse fitting~(see Supplementary Information for details).
While this approach is numerically stable, non-iterative,
and guarantees an ellipse-specific solution,
it doesn't work well at $\phi_d$ closes to 0 or $\pi$ \cite{young_half_minute_scale_2020}.
Moreover,
the effective probability distribution the data is sampled from is the convolution of an ellipse and a binomial distribution associated with QPN.
Therefore,
the bias error needs to be accounted for in order to extract the correct differential frequencies between ensemble pairs.
To do this,
we perform Monte-Carlo simulations which generate artificial data with contrast and atom number for each ensemble that captures the QPN,
and known differential phases as input parameters.
The simulated data allows us to calculate a correction phase with a statistical standard deviation as the error bar in the bias correction.

To illustrate the validity and importance of bias correction,
Extended Data Fig.\ref{figED.Loop_result_compare} shows the comparison between ``closed-loop'' analysis within 6 ensembles (Fig.~\ref{fig4}\textbf{c}) with (filled red points) and without (empty blue points) bias correction.
The sum frequencies of each unique ``closed-loop'' agree within an $1\times10^{-18}$ window with bias correction,
while the deviations from zero are as large as $1\times10^{-17}$ without bias correction.
We note that while this bias can be avoided through a judicious choice of phase for 2 clocks, it is unavoidable in differential clock comparisons with 3 or more clocks.
For example, in the extreme case where two pairs of clocks are operating at differential phases of multiples of $\pi/2$ where the bias error is minimized,
i.e. $\phi_{12}=(2m+1)\pi/2$ and $\phi_{23} = (2n+1)\pi/2$,
where $m, n$ are integers.
The outcome of the third pair would be $\phi_{13}= \phi_{12} + \phi_{23} = ((m+n) +1) \pi$,
which is a multiple of $\pi$ where the bias error is maximized.

\subsection*{\hfil Units and errors \hfil}
Unless otherwise stated, 
all errors and numerical uncertainties in this article and its Supplementary Information denote a 1$\sigma$ s.d. confidence interval. 
When we quote a coherence time,
we are typically referring to the 1/e decay time.
When we explicitly refer to a Gaussian time constant, 
we are referring to the timescale associated with 1 s.d. of the Gaussian envelope.

\subsection*{\hfil Data and code availability \hfil} 
The experimental data presented in this manuscript and the code used for analysis and simulation in this work are available from the corresponding author upon reasonable request. 

\subsection*{\hfil Acknowledgements \hfil} We thank Jun Ye, Adam Kaufman, Jeff Thompson, Toby Bothwell, and Andrew Jayich for insightful discussions and helpful feedback on the manuscript.  This work was supported in part by the NIST Precision Measurement Grants program, the Northwestern University Center for Fundamental Physics and the John Templeton Foundation through a Fundamental Physics grant, the Wisconsin Alumni Research Foundation, the Army Research Office through agreement number W911NF-21-1-0012, and a Packard Fellowship for Science and Engineering.

\subsection*{\hfil Author contributions \hfil} X.Z.~designed and built the experimental apparatus with assistance from J.D., H.L., and B.M., and with guidance from S.K. All authors contributed to maintenance and operation of the experimental apparatus, data collection, data analysis, and to writing the manuscript.

\subsection*{\hfil Competing interests \hfil}
The authors declare no competing interests.

\pagebreak
\section*{\hfil Extended Data \hfil}

\captionsetup[figure]{labelfont={bf},name={Extended Data Figure}, labelsep=period}
 
\renewcommand{\thefigure}{1}
\begin{figure}[!ht]
    \centering
    \includegraphics[width=0.75\textwidth]{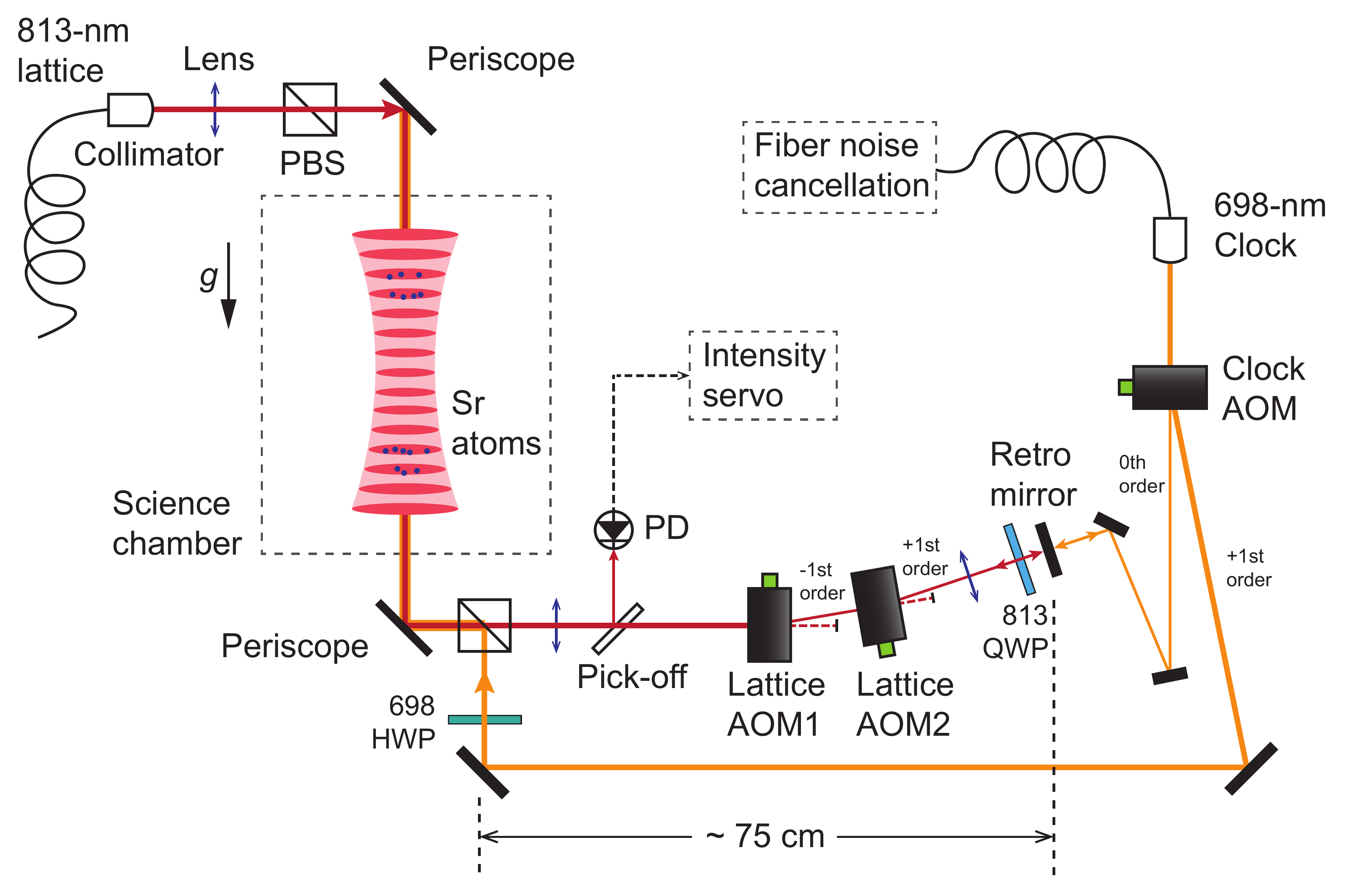}
    \caption{\textbf{Lattice and clock path.}
    Schematic diagram showing the lattice and clock beam paths for the interrogation the $\ket{{}^1S_0, m_F=\pm5/2}\leftrightarrow \ket{{}^3P_0, m_F=\pm3/2}\sigma$-transition. 
    To interrogate the $\ket{{}^1S_0, m_F=\pm9/2}\leftrightarrow \ket{{}^3P_0, m_F=\pm9/2}\sigma$-transition, 
    the first order diffraction clock beam is overlapped with the lattice using a long-pass dichroic beam-splitter,
    which is not shown in this figure.
    PBS: polarized beam-splitter;
    AOM: acousto-optic-modulator;
    PD: photo-diode; 
    Sr: strontium; 
    HWP: half-waveplate;
    QWP: quarter-waveplate.
    }
    \label{figED.clock_lattice_path}
\end{figure}

\pagebreak

\renewcommand{\thefigure}{2}
\begin{figure}[!ht]
    \centering
    \includegraphics[width=0.75\textwidth]{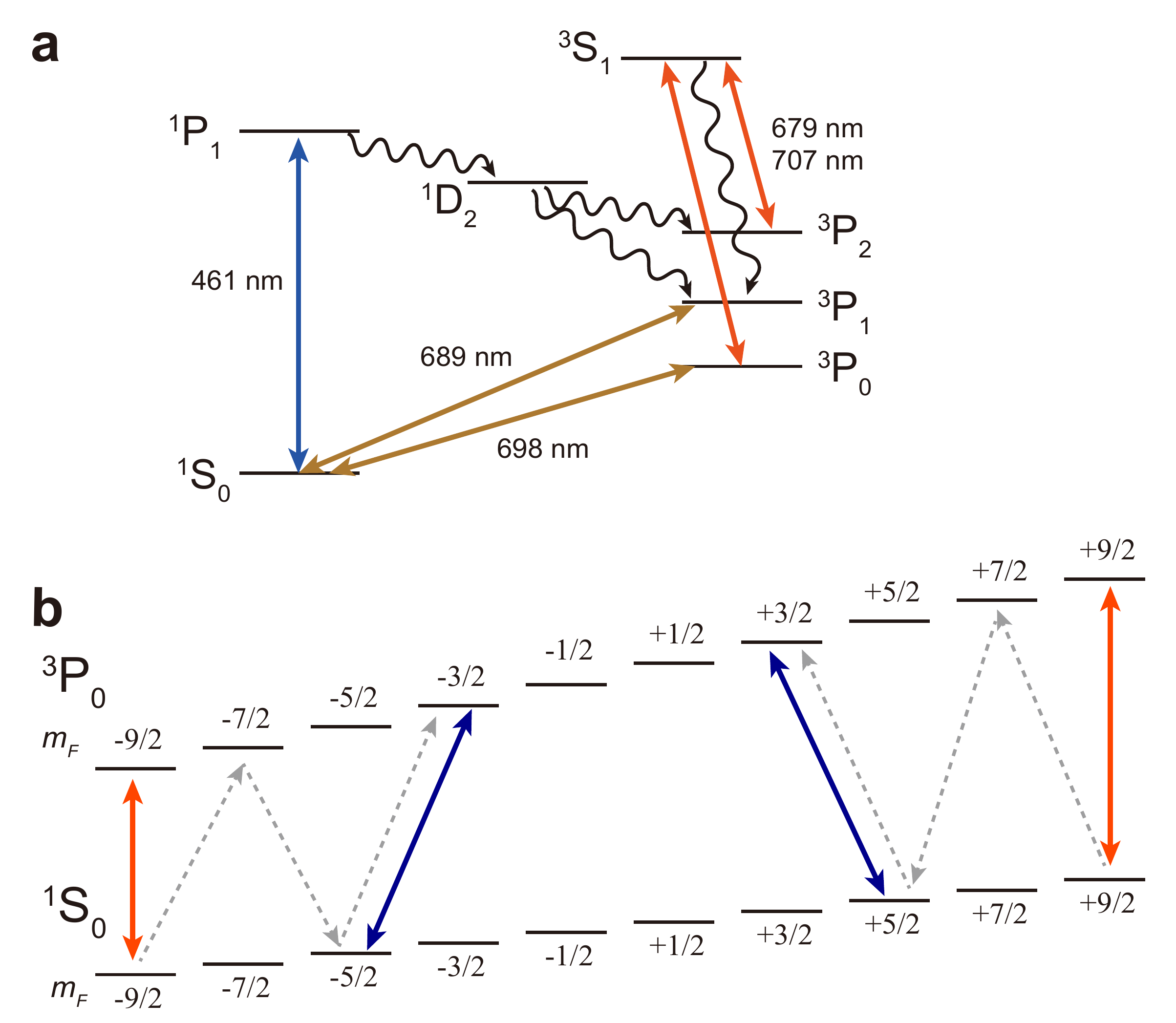}
    \caption{\textbf{Energy levels for strontium and hyperfine states for clock interrogation.}
    (a) Energy level diagram for strontium. ${}^1S_0 \leftrightarrow {}^1P_1$ transition on 461 nm for first-stage MOT and imaging. ${}^1S_0 \leftrightarrow {}^3P_1$ transition on 689 nm for second-stage MOT.
    ${}^3P_2 \leftrightarrow {}^3S_1$ transition on 679 nm and ${}^3P_0 \leftrightarrow {}^3S_1$ transition on 707 nm for repumping.
    ${}^1S_0 \leftrightarrow {}^3P_0$ transition on 698 nm for clock spectroscopy.
    (b) 10 hyperfine states for clock interrogation. 
    Red double arrows represent clock interrogation on the $\ket{{}^1S_0, m_F=\pm9/2}\leftrightarrow\ket{{}^3P_0, m_F=\pm9/2}$ transition. 
    Blue double arrows represent clock interrogation on the $\ket{{}^1S_0, m_F=\pm5/2}\leftrightarrow\ket{{}^3P_0, m_F=\pm3/2}$ transition. 
    Grey dashed lines stand for transitions for coherent transfer of atoms from $m_F=\pm9/2$ states to $m_F=\pm3/2$ states.}
    \label{figED.energy_levels}
\end{figure}

\pagebreak

\renewcommand{\thefigure}{3}
\begin{figure}[!ht]
    \centering
    \includegraphics[width=0.95\textwidth]{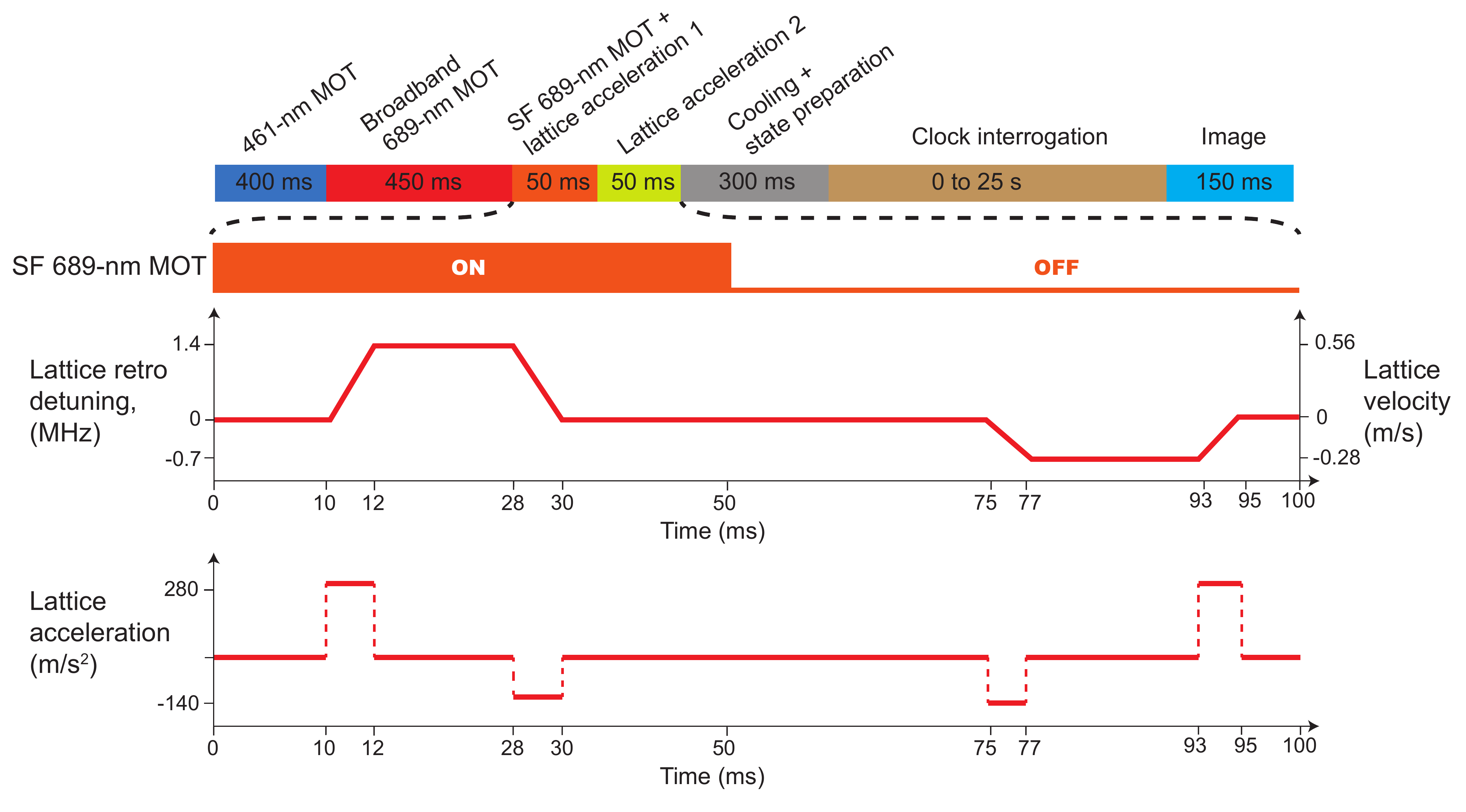}
    \caption{A typical timing diagram for a Ramsey spectroscopy sequence,
    in which laser cooling,
    state preparation and camera imaging contribute to about 1.6 s dead time,
    with clock interrogation time ranging from 10 ms to 25 s. 
    The two plots below are the corresponding lattice retro detuning,
    lattice velocity and lattice acceleration during loading two ensembles at 1 cm separation.}
    \label{figED.timing_diagram}
\end{figure}

\pagebreak

\renewcommand{\thefigure}{4}
\begin{figure}[!ht]
    \centering
    \includegraphics[width=0.95\textwidth]{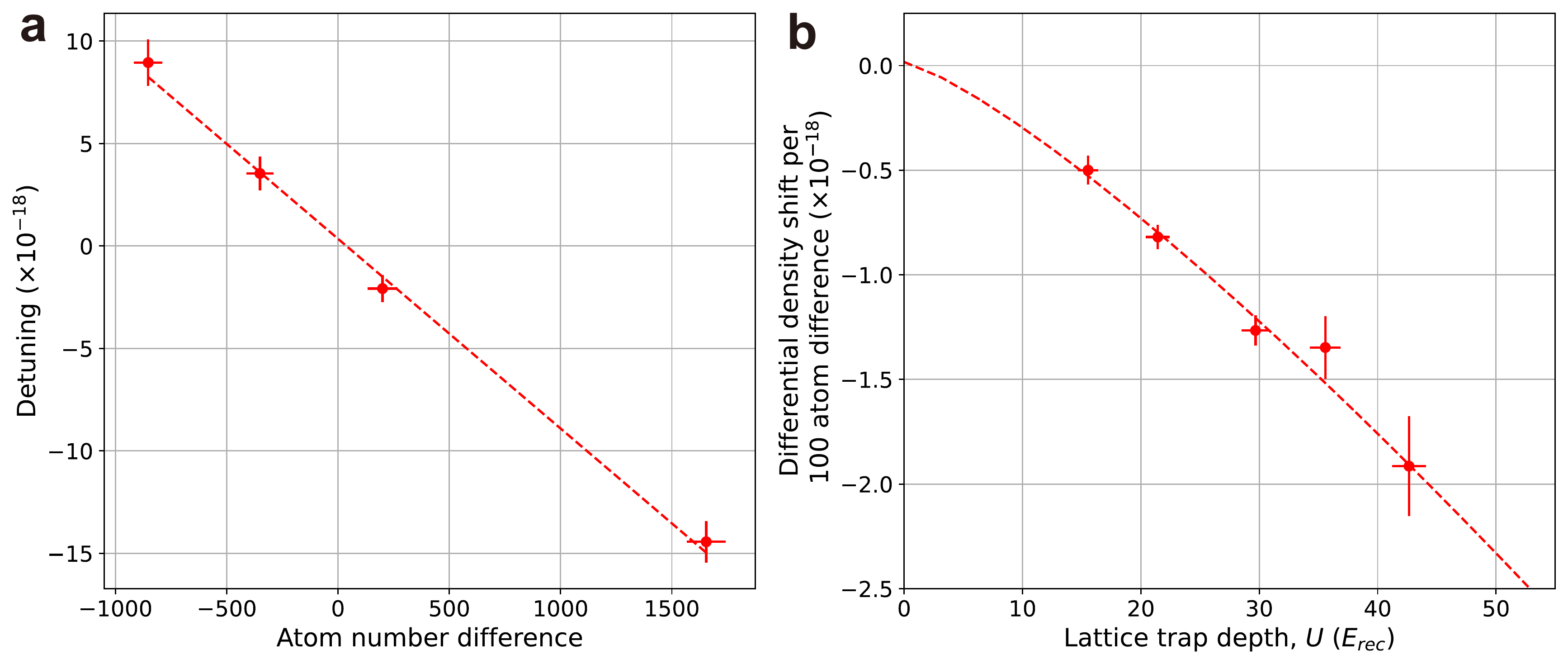}
    \caption{(a) Differential density shift as a function of relative atom number difference between two symmetrically prepared ensembles at 1 cm separation. The data is taken at  20 $E_{\textrm{rec}}$ lattice trap depth. Dashed line is the linear fitting, in which the slope is extracted as $-8.5(6)\times10^{-19}$ shift per 100 atom number difference.
    (b) Scaling of differential density shift per 100 atom number difference between ensemble pairs with lattice trap depth $U$.
    The dashed line is a fit to the expected $\alpha U^{5/4} + \beta$ scaling,
    where $\alpha$ and $\beta$ are fit parameters.
    }
    \label{figED.DensityShifts}
\end{figure}

\pagebreak

\renewcommand{\thefigure}{5}
\begin{figure}[!ht]
    \centering
    \includegraphics[width=0.95\textwidth]{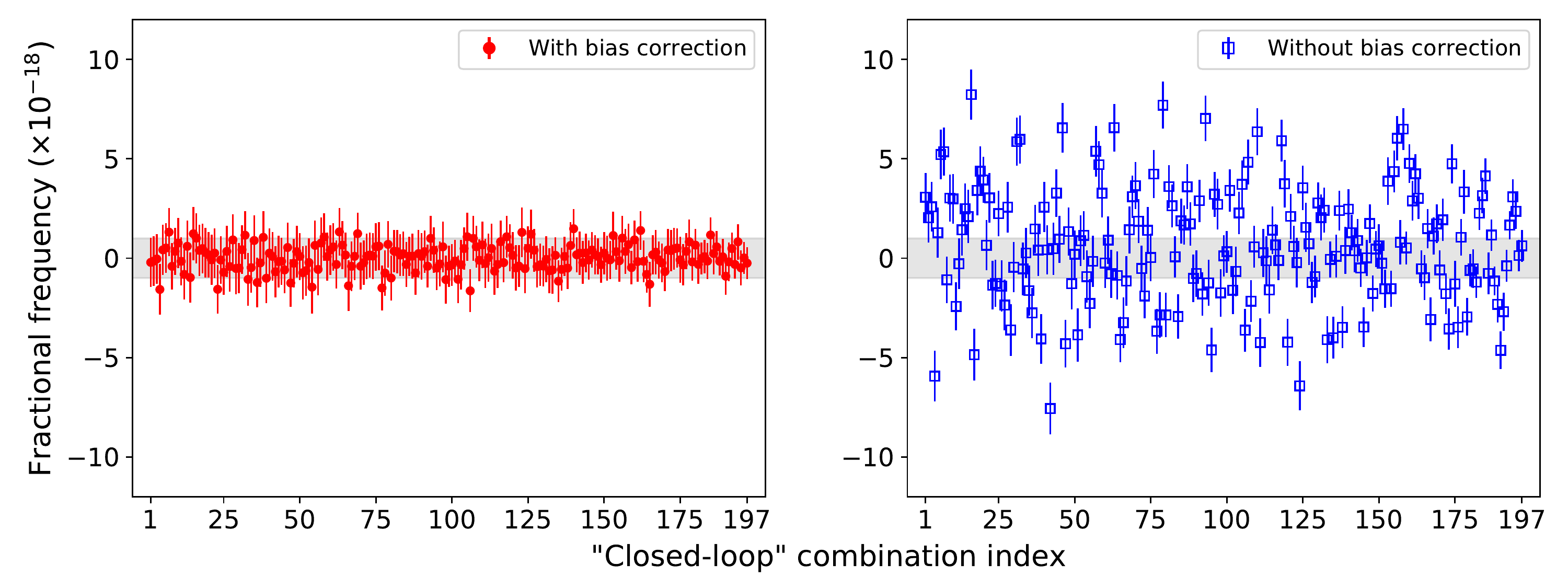}
    \caption{Comparison of ``closed-loop'' analysis with (left panel, red points) and without (right panel, blue points) bias correction.
    197 unique ``closed-loop'' combinations are shown, with each datum corresponds to the sum frequency within each loop.
    Shaded area represents an $1\times10^{-18}$ window.
    }
    \label{figED.Loop_result_compare}
\end{figure}

\pagebreak

\captionsetup[table]{labelfont={bf},name={Extended Data Table},
 labelsep=period}

\renewcommand{\thetable}{1}
\begin{table}[!ht]
\centering
\begin{tabular}{ |p{6.5cm}||p{3.0cm}|p{3.0cm}| }
 \hline
 Approaches  &  $1/e$ decay time & Fitting \\
 \hline
 Ramsey with one ensemble  &    96(24) ms  &     Gaussian   \\
 \hline
 Ramsey with 2 ensembles, $9/2-9/2$  &    6(1) s  &     Gaussian   \\
 \hline
 Spin-echo, $9/2-9/2$  &  24(5) s  & exponential  \\
 \hline
 Ramsey with 2 ensembles, $5/2-3/2$ &  26(2) s  & exponential  \\
 \hline
\end{tabular}
\caption{\label{tab:contrast} Table of measured coherence times using different approaches. Uncertainties are quoted as 1$\sigma$ standard deviations.}
\end{table}

\pagebreak

\renewcommand{\thetable}{2}
\begin{table}[!ht]
    \centering
    \begin{tabular}{|l|c|c|c|c|c|c|}\hline
    \diaghead{\theadfont Diaddddds}{$j$}{$i$}&
    1& 2 & 3 & 4 & 5 & 6\\
    \hline
    1 & --- & --- & --- &---  &--- & --- \\
    \hline
    2 & 61.17(16) & --- & --- &---  &--- & --- \\
    \hline
    3 & 125.74(19) & 64.67(17) & --- &---  &--- & --- \\
    \hline
    4 & 196.67(15) & 135.57(15) & 70.65(18) &---  &--- & --- \\
    \hline
    5 & 274.83(15) & 213.69(22) & 148.88(23) & 78.15(20) &--- & --- \\
    \hline
    6 & 360.19(15) & 299.02(20) & 233.96(23) &  163.66(16)  &  85.43(25) & --- \\
    \hline
    \end{tabular}
    \caption{Table of differential frequencies ($\delta f_{ji}=-\delta f_{ij}=f_j - f_i$, where $i,j$ are indices of ensembles) from 6 ensemble, 15 pairwise comparisons. All units are in mHz, with error bars of 1$\sigma$ standard deviation.}
    \label{15_pairwise_results}
\end{table}

\pagebreak

\section*{\hfil Supplementary Information \hfil}

\renewcommand{\contentsname}{\vspace{-\baselineskip}} % remove Content title

\tableofcontents

\pagebreak

\captionsetup[figure]{labelfont={bf},name={Figure S}, labelsep=period}

\subsection{Balanced trap depths of symmetrically prepared ensembles}

To verify that the ensemble pairs are symmetrically prepared relative to the lattice beam waist and have consistent atomic temperatures,
we perform motional sideband spectroscopy on the $\ket{^1S_0}\leftrightarrow\ket{^3P_0}$ transition with a pulse duration of 150 ms (see Fig~S.~\ref{figS.BalancedDepth}).
The red (blue) sideband corresponds to transition from $\ket{g, n_g}\leftrightarrow \ket{e, n_e = n_g-1 }$ ($\ket{g, n_g}\leftrightarrow \ket{e, n_e = n_g+1 }$),
where $n_{g,e}$ is the vibrational quantum number in the ground (excited) state.
The lattice trap depth is determined by the cut-off frequency of the sidebands~\cite{blatt_rabi_2009},
and the ratio of the area under the blue and red sidebands is used to extract the axial temperature.
The lattice alignment is optimized such that the trapping frequencies agree within 1-kHz resolution,
which is equivalent to a trap depth difference of below $1~E_{\textrm{rec}}$ for typical depths of $\sim20~E_{\textrm{rec}}$.
The temperatures of the two ensembles agree within $0.1~\mu$K,
inferred by the axial temperature ($0.7~\mu$K and mean quantum occupation number of 0.15) extracted from the motional sidebands and the radial temperature ($\sim0.6~\mu$K) is determined by probing the Doppler broadened profile with a separate clock beam path orthogonal to the optical lattice.

\renewcommand{\thefigure}{1}
\begin{figure}[!ht]
    \centering
    \includegraphics[width=0.75\textwidth]{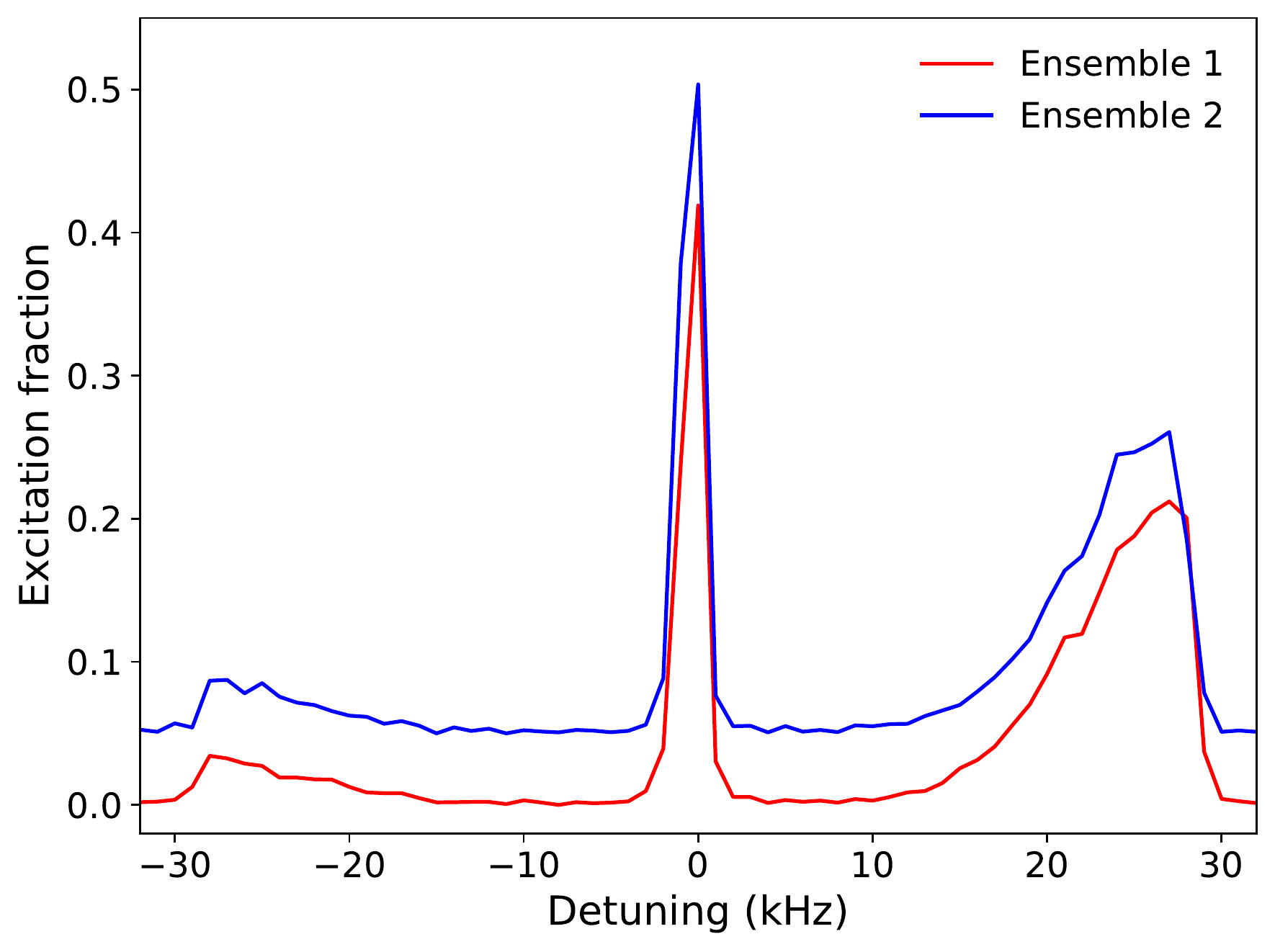}
    \caption{Motional sideband spectroscopy for two ensembles separated by 1 cm with trap depth of 20 $E_{\textrm{rec}}$ taken with a 150 ms pulse on the $^1S_0\leftrightarrow{}^3P_0$ transition. The fitted sidebands give the mean vibrational quantum state occupation number and axial temperature, 
    and are consistent for both ensembles within the 1-kHz resolution.
    The excitation fractions of ensemble 2 is shifted up by +0.05 along the y-axis for clarity.}
    \label{figS.BalancedDepth}
\end{figure}

\subsection{Raman scattering, lifetime and lattice trap depth dependent contrast}

To reduce the off-resonant lattice photon induced Raman scattering~\cite{dorscher_lattice_induced_2018},
one would prefer to operate at shallower lattice trap depths.
While gravity creates a potential energy difference between adjacent lattice sites and suppresses tunneling for vertical 1D-lattices,
we observe reduced lifetimes for both ground and excited state atoms at shallower trap depths,
likely due to residual parametric heating from the lattice.
At deeper trap depths ($>30~E_{\textrm{rec}}$),
the lifetimes for atoms in the excited clock state drops below 15 seconds,
which is likely limited by Raman scattering from the lattice light.
However,
we also observe a decrease in the ground state lifetimes at much deeper lattice trap depths,
suggesting lattice-intensity-dependent heating.

The measured Ramsey contrasts for ensembles prepared with approximately 2000 atoms start to drop below 0.6 as the lattice trap depths increase from 20 $E_{\textrm{rec}}$  (See FigS.~\ref{figS.Lifetime}).
This is likely due to a combination of Raman scattering,
which scales linearly to lattice trap depth $U$,
and atomic density which scales as $U^{5/4}$.
The competition between reduced lifetime,
reduced atom density and increased contrast when lowering the lattice trap depth leads to an optimal trap depth between 15 to 20 $E_{\textrm{rec}}$ for Ramsey interrogation under our current operating conditions.
For the experiments shown in the main text,
we choose to operate at a lattice trap depth of 20 $E_{\textrm{rec}}$,
at which we measure a 24(3) s lifetime for ground state atoms, a
13(2) s lifetime for excited state atoms,
and Ramsey contrast of 0.65 for 2400 atoms per ensemble.

\renewcommand{\thefigure}{2}
\begin{figure}[!ht]
    \centering
    \includegraphics[width=0.75\textwidth]{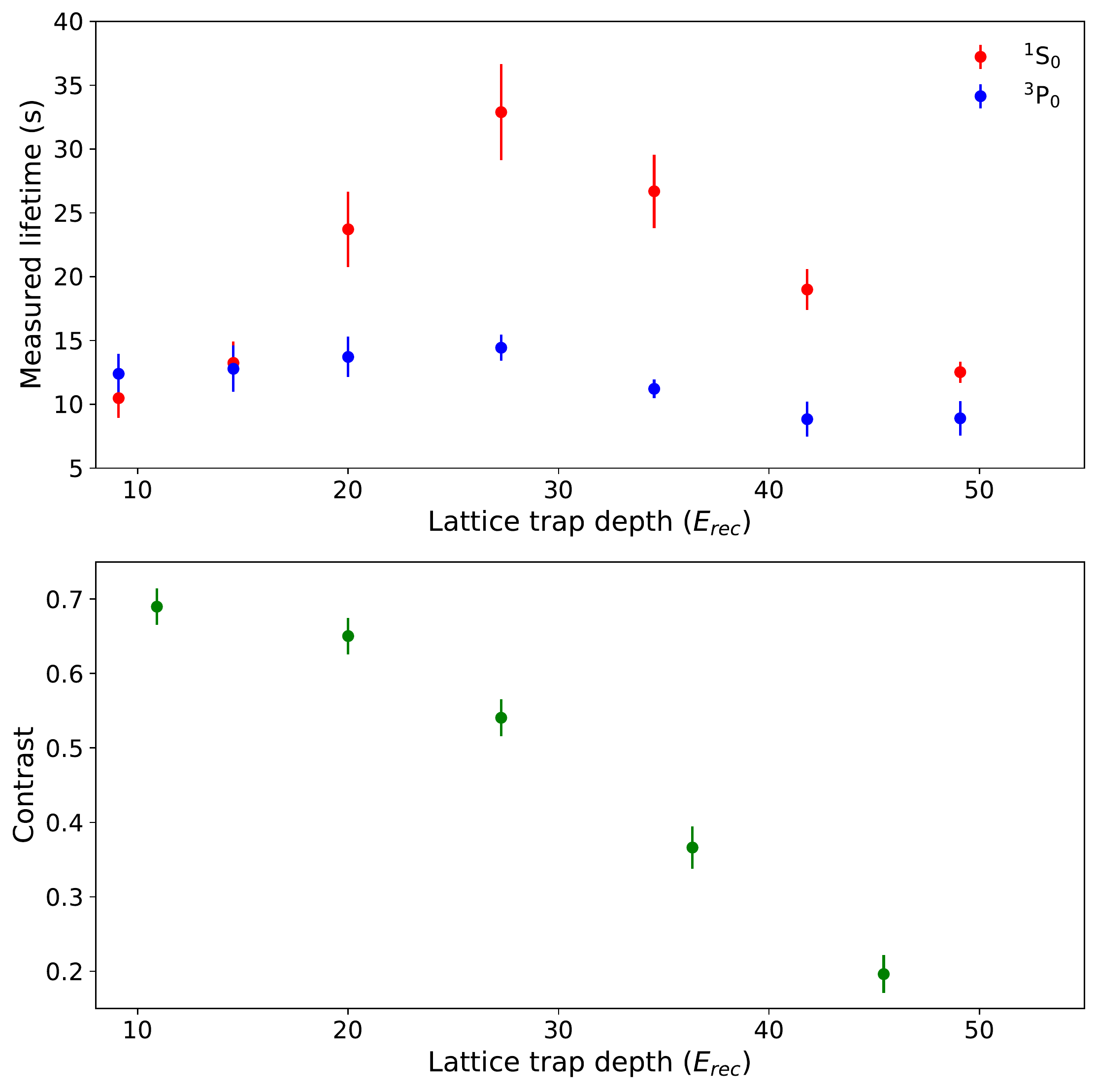}
    \caption{Top: Measurement of ground (red) and excited (blue) state atom lifetimes as a function of lattice trap depth. 
    Bottom: Measurement of Ramsey contrast at 8 s dark time and approximately 2000 atoms per ensemble at different lattice trap depths.}
    \label{figS.Lifetime}
\end{figure}

\subsection{Operational magic wavelength}

The ac Stark shift at a particular trap depth $U_0$ for $\pi$ ($\Delta m_F = 0$) transition starting from ground state $m_F$ can be expressed as~\cite{boyd_m._high_2007, Westergaard_acStark_2011}
\begin{equation}
    \Delta\nu_{\textrm{ac}} = \bigg(\Delta\kappa^{S} + \Delta\kappa^{V} m_F \xi \hat{\epsilon}_k\cdot\hat{\epsilon}_B  + \Delta\kappa^{T} (3\textrm{cos}^2\theta -1 )\big[3m_F^2 - F(F+1)\big]\bigg) U_0, 
\end{equation}
where $\Delta\kappa^{S,V,T}$ are the differential scalar, vector, and tensor shift coefficients between excited and ground states, respectively;
$\xi$ is the lattice light ellipticity;
$\hat{\epsilon}_{k, B}$ are units vectors along the lattice beam wave vector and magnetic field quantization axis, respectively;
$\theta$ is the angle between the (nearly) linear lattice polarization and $\hat{\epsilon}_B$.

Similarly, 
the ac Stark shift for $\sigma^\pm$ ($\Delta m_F = \pm1$) transition starting from ground state $m_F$ has the form
\begin{equation}
\begin{split}
    \Delta\nu_{\textrm{ac}} =  & \bigg(\Delta\kappa^{S} + \big(\kappa_e^{V} (m_F\pm1) - \kappa^V_g m_F \big) \xi \hat{\epsilon}_k\cdot\hat{\epsilon}_B  \\
    & +\Delta\kappa^{T} (3\textrm{cos}^2\theta -1 )\big[- F(F+1)\big]\\
    & +(3\textrm{cos}^2\theta -1)\big[ \kappa_e^T(3m_F\pm1)^2 - \kappa^T_g 3m_F^2 \big]\bigg) U_0,
\end{split}
\end{equation}
where $\kappa_{g}^{V,T}$ ($\kappa_{e}^{V,T}$) are vector, tensor coefficients for ground (excited) states,
respectively.

An operational magic wavelength corresponds to the lattice frequency where the scalar and tensor Stark shifts cancel,
and the remaining vector Stark shift can be eliminated by both using a linearly polarized lattice beam ($\xi\approx0$) and averaging between the $\pm m_F$ manifolds.
We note that the above equation doesn't account for higher order effects such as hyperpolarizability~\cite{brown_hyperpolarizability_2017,Ushijima_operational_2018},
which scales quadratic to the lattice trap depth,
and is negligible in differential clock comparison between two ensembles at shallow lattice trap depths.
For example,
at $20~E{\textrm{rec}}$ trap depths,
assuming a $1~E{\textrm{rec}}$ trap depth difference,
the hyperpolarizability induces a differential ac Stark shift of less than $8\times10^{-20}$
\cite{bothwell_jila_2019_SM}.

In this work,
the operational magic wavelength is chosen to be 368554.4849(1) GHz for $\ket{^1S_0,m_F = \pm9/2}$ $\leftrightarrow\ket{^3P_0, m_F = \pm9/2}$ $(\pi$ transition),
which is given by the previous experiment~\cite{nicholson_systematic_2015}.
However,
the above wavelength no longer works for the $\ket{^1S_0,m_F = \pm5/2}\leftrightarrow\ket{^3P_0, m_F = \pm3/2}$~$(\sigma$ transition) because of $m_F$ dependence in tensor Stark shift.
In the limit where the lattice frequency is near the magic wavelength,
we have $\Delta \kappa ^T = \kappa^T_e - \kappa^T_g \simeq \kappa^T_e = -0.0058(23)$~mHz$/E_{\textrm{rec}}$~\cite{Westergaard_acStark_2011}.
We would expect a differential shift of $-117~\Delta\kappa^T U_0$ for $\ket{g, \pm5/2}\leftrightarrow\ket{e, \pm3/2}$ transition using the above operational magic wavelength and assuming $\theta \approx 0$.
At a typical trap depth of $U_0 = 20~E_{\textrm{rec}}$,
this corresponds to a shift of $\approx+13.5$ mHz.

To find the operational magic wavelength for $\ket{g, \pm5/2}\leftrightarrow\ket{e, \pm3/2}$ transition,
the lattice frequency is scanned across a range of $\pm 800$ MHz and the contrasts of synchronized Ramsey interrogations are measured at each frequency (main text Fig.2e).
The optimal contrast is found at a lattice frequency of 368554.810(30) GHz,
which is blue shifted by $+325(30)$ MHz compared to the operational magic wavelength for $\ket{g, \pm9/2}\leftrightarrow\ket{e, \pm9/2}$ transition.

\subsection{Differential Zeeman shifts and magnetic field sensitivities}

For the $\ket{{}^1S_0\leftrightarrow{}^3P_0}$ clock transition,
the linear Zeeman shift at a magnetic field $B$ for $\pi$ transition starting from a ground hyperfine state $m_F$ can be written as~\cite{boyd_nuclear_2007}
\begin{equation}
    \Delta\nu_{L, \pi} = -\delta g m_F \mu_0 B,
\end{equation}
where $\delta g$ is the differential landé $g$-factor between ground and excited states.
$\mu_0 = \mu_B /h$,
in which $\mu_B$ is the Bohr magneton and $h$ is the Planck constant.

Similarly,
we can express the linear Zeeman shift for $\sigma^\pm$ transition from a ground state $m_F$ as
\begin{equation}
    \Delta\nu_{L, \sigma^\pm} = -(\pm g_I + \delta g( m_F\pm 1) )\mu_0 B,
\end{equation}
where $g_I$ is the nuclear landé $g$-factor.

With $\delta g \mu_0 = -108.4$ Hz/G and $g_I \mu_0 = -185 $ Hz/G as input~\cite{boyd_nuclear_2007},
we would expect linear Zeeman shifts for $\ket{g, \pm9/2}\leftrightarrow\ket{e, \pm9/2}$~($\pi$ transition)
\begin{equation}
    \Delta_{ZS}^{\pm9/2\leftrightarrow\pm9/2,\pi} = \pm 487.8 ~\textrm{Hz/G},
\end{equation}
and similarly for for $\ket{g, \mp5/2}\leftrightarrow\ket{e, \mp3/2}$~($\sigma^\pm$ transition)
\begin{equation}
    \Delta_{ZS}^{\mp5/2\leftrightarrow\mp3/2,\sigma^{\pm}} = \pm 22.4 ~\textrm{Hz/G},
\end{equation}
which is a factor of 22 smaller than that of $\ket{g, 9/2}\leftrightarrow\ket{e, 9/2}$~($\pi$ transition).

The quadratic Zeeman shift has negligible $m_F$ dependence and can be written as
\begin{equation}
    \Delta\nu_{Q} = \delta_B^{(2)}\mu_0 B^2,
\end{equation}
where $\delta_B^{(2)}\mu_0 = -0.233(5)$ Hz/G$^2$ is the quadratic Zeeman shift coefficient.

Under a typical bias magnetic field of 2 G and a magnetic field gradient of 15 mG/cm,
the differential linear Zeeman shift (for $\sigma^-$ transition, $\ket{g, +5/2}\leftrightarrow\ket{e, +3/2}$) between two ensembles separated by 1 cm is approximately 350 mHz,
and the differential quadratic Zeeman shift is approximately 14 mHz.

\subsection{Ellipse fitting}
In order to extract the differential frequency detuning between the two ensembles we interrogate,
we follow the procedure demonstrated by~\cite{marti_imaging_2018, young_half_minute_scale_2020}.
After a Ramsey dark time $\tau_d$,
the excitation fraction of each ensemble can be expressed as 
\begin{equation}
    P_{1} = \frac{1}{2}\bigg(1+C\big[\text{cos}(\omega_{1}-\omega_{l})\tau_d\big]\bigg),
\end{equation}
\begin{equation}
    P_{2} = \frac{1}{2}\bigg(1+C\text{cos}\big[(\omega_{1} + \omega_{d}-\omega_{l})\tau_d\big]\bigg),
\end{equation}
where $C$ is the contrast,
$\omega_{1}$ is the frequency of the ensemble 1,
$\omega_{l}$ is the frequency of the laser, 
and $\omega_{d}$ is the frequency difference between ensemble 2 and 1.
We can then re-express these excitation fractions as functions of angles $\theta$ and $\phi$, where $\theta$ is the atom-laser phase, $(\omega_{0}-\omega_{1})\tau_d$,
and $\phi$ is the differential phase between regions, $\omega_{d}t$. 
\begin{equation}
    P_{1} = \frac{1}{2}\bigg(1+C\text{cos}(\theta)\bigg),
\end{equation}
\begin{equation}
    P_{2} = \frac{1}{2}\bigg(1+C\text{cos}(\theta + \phi)\bigg).
\end{equation}

Since we are operating at Ramsey interrogation times well beyond the laser coherence time, $\theta$ is random for each experiment and is uniformly distributed from $0$ to $2\pi$, while $\phi$ stays constant across experiments.

In order to extract $\phi$ from our data, 
we plot the excitation fraction in each ensemble for a given shot as a single point on a parametric plot, 
with ensemble 2 on the vertical axis and ensemble 1 on the horizontal axis. 
As shots build up,
an ellipse is traced out,
with points randomly sampling the perimeter of the ellipse due to the random distribution of $\theta$. 
We then fit to this ellipse using least-squares approach~\cite{LSQ_ellipse_1998} and extract $\phi$ through $\phi$ = 2arctan(b/a),
where $a$ and $b$ are the extracted semi-major and semi-minor axis,
respectively.

\subsubsection{Phase extraction variance and biased error}
In order to accurately determine our uncertainty in extracting $\phi$,
we calculate the variance of $\phi$ through the variance in $P_{1}$ and $P_{2}$ due to QPN. 
For convenience, we can define
\begin{gather}
    x = \frac{C}{2} \text{cos}(\theta), \\
    y = \frac{C}{2} \text{cos}(\theta + \phi),
\end{gather}
such that we can express the variance of $\phi$ as 
\begin{equation}
    \sigma^{2}(\phi) = \abs{\frac{\partial \phi}{\partial x}}^{2} \sigma^{2}(x) + \abs{\frac{\partial \phi}{\partial y}}^{2} \sigma^{2}(y),
\end{equation}

The partial derivatives can be evaluated through Jacobian matrix inversion, and the variance in x and y due to quantum projection noise (QPN) can be expressed as 
\begin{gather}
    \sigma^{2}(x) = \frac{1}{N_{1}}P_{1}(1-P_{1}), \\
    \sigma^{2}(y) = \frac{1}{N_{2}}P_{2}(1-P_{2}),
\end{gather}
which gives a expression for the variance of $\phi$
\begin{equation}
    \sigma^{2}(\phi) = \frac{4}{C^2} \bigg(\text{csc}^2(\theta) \sigma^{2}(x) + \text{csc}^{2}(\theta+\phi)\sigma^{2}(y)\bigg)
\end{equation}
Finally, since we take repeated measurements of $\phi$ for a random $\theta$, we average over a uniform $\theta$ distribution to get an average variance in $\phi$ as the following.
\begin{equation}
    \langle \sigma^{2}(\phi) \rangle = \frac{4}{C^2} \bigg( \int_{0}^{2\pi} \frac{d\theta}{2\pi} \frac{1}{\text{csc}^2(\theta) \sigma^{2}(x) + \text{csc}^{2}(\theta+\phi)\sigma^{2}(y)}     \bigg)^{-1}
    \label{qpn_limit_ellipse}
\end{equation}

\renewcommand{\thefigure}{3}
\begin{figure}[!ht]
    \centering
    \includegraphics[width=0.75\textwidth]{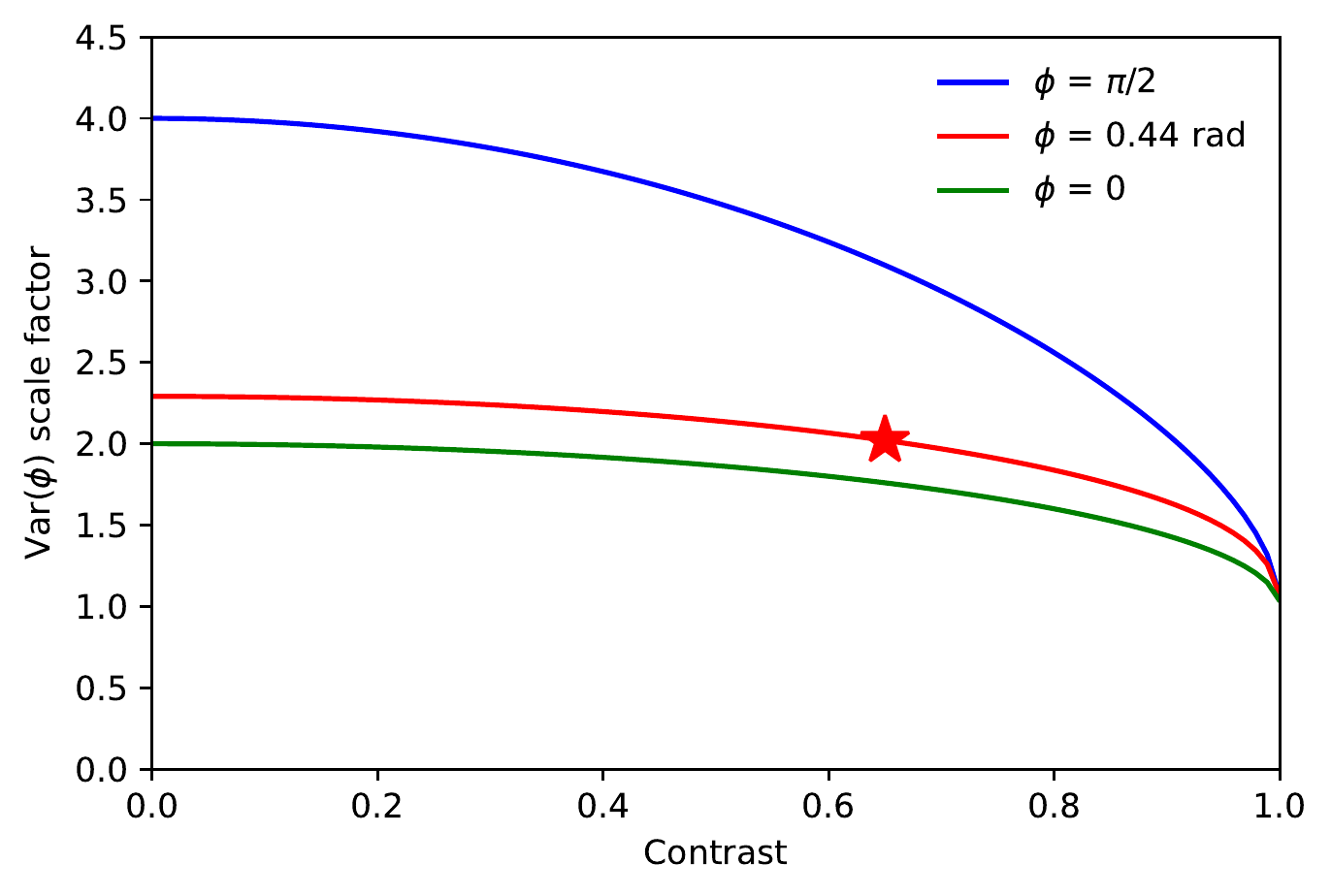}
    \caption{Additional scale factor for variance of differential phase $\phi$. The red star corresponds to the measurement of relative stability taken in the main text (Fig.\ref{fig3}) with $\phi\approx 0.44$ rad and $C\approx0.65$.
    }
    \label{figS.QPN_scale_factor}
\end{figure}

In the case where $C$ = 1,
QPN results in a variance of $\langle \sigma^{2}(\phi) \rangle = 2/(NC^2)$,
which is the familiar result for a QPN-limited Ramsey spectroscopy.
For $C<1$,
Var($\phi$) is scaled by a factor (see Fig~S~\ref{figS.QPN_scale_factor}),
which depends on the phase and contrast.
Counter-intuitively,
the QPN limit is minimized at $\phi \approx 0$ or $\pi$ (a line) where the fits are biased,
and maximized at $\pi/2$ where it's a circle with least biased errors.
This would suggest operating at a differential phase $\phi$ closer to 0 or $\pi$ to get lower QPN,
while removal of biased errors should be considered.
This can be quantified by running Monte-Carlo simulations with known phases and experimental parameters as input,
that we can bound the biased error in ellipse fitting below 3\% at $\phi\approx$ 0.44 rad (see Fig~S~\ref{figS.BiasedError}).

\renewcommand{\thefigure}{4}
\begin{figure}[!ht]
    \centering
    \includegraphics[width=0.75\textwidth]{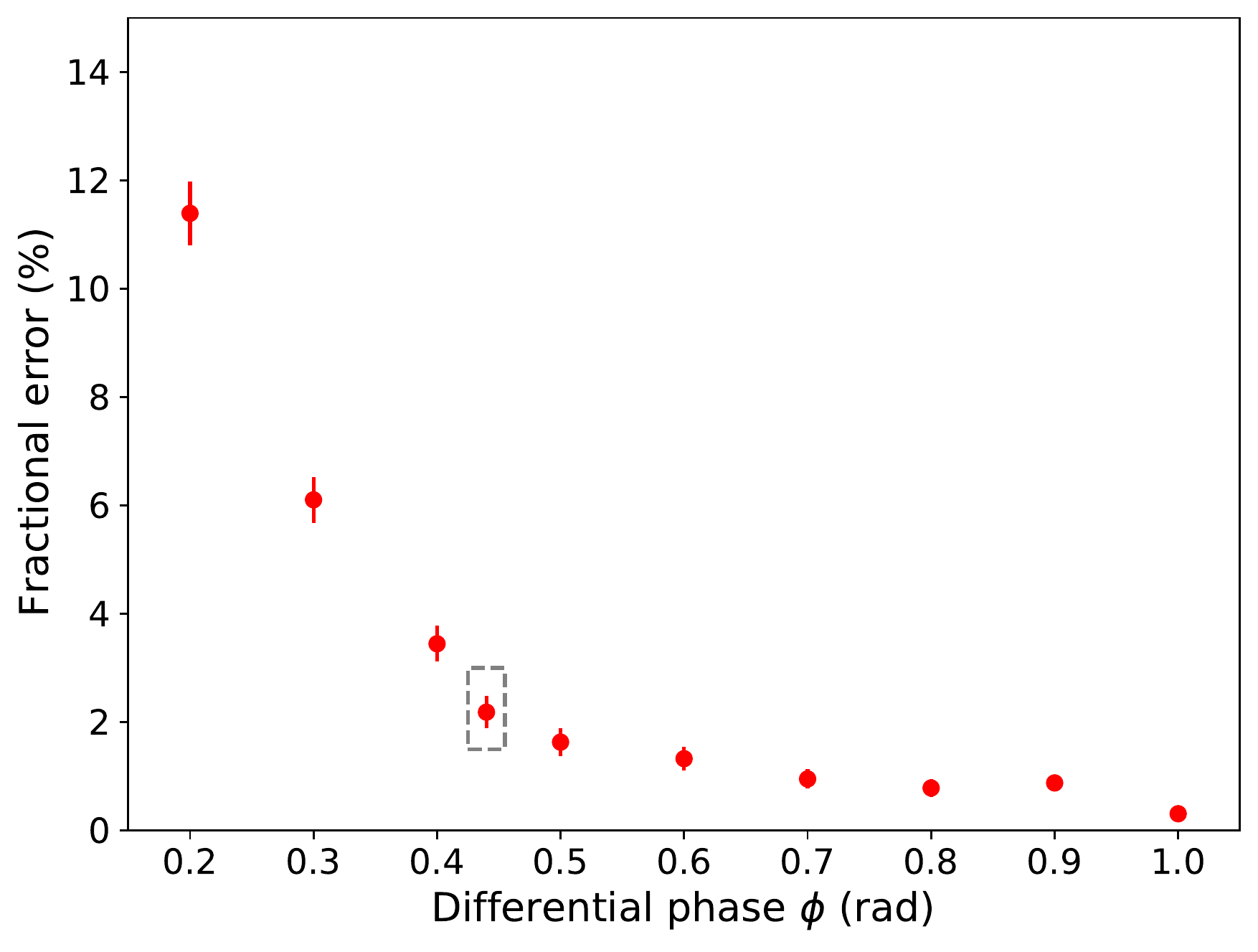}
    \caption{Monte-Carlo simulations of fractional biased error in ellipse fitting for 2400 atoms in each ensemble. The grey box corresponds to $\phi\approx$ 0.44 rad,
    at which the biased error can be bound below 3\%.
    }
    \label{figS.BiasedError}
\end{figure}

\subsubsection{Determination of Ramsey contrast}
%It is not entirely straight-forward to determine the contrast for each ensemble from the ellipse fitting independently,
%especially in the extreme case that $\phi = 0$ or $\pi$,
%where the ellipse fitting is heavily biased.
%The difference between the maximal and the minimal excitation fractions does not yield the correct contrast in the presence of QPN.
%Instead,
%we use an alternative approach,
%which is independent of the ellipse fitting and the offset phase $\phi$,
%to determine the ensemble contrast independently.
We determine the Ramsey contrast independently for each ensemble, rather than extracting it from the fitted ellipse.
To do so, we plot the histogram of the excitation fractions of each ensemble which follows a bimodal distribution,
and the contrast is subtracted by mapping the two local maximums.
This is then corrected for a small offset based on the Monte-Carlo simulations with known contrasts and QPN as input.
Fig~S~\ref{figS.ContrastMC}a shows a simulated ellipse and its fitting at $\phi = 0$ rad with 500 atoms, 100 measurement runs and 0.65 contrast for each ensemble.
The corresponding histograms for excitation fractions are shown on the top and right axes.
Fig~S~\ref{figS.ContrastMC}b) is the Monte-Carlo simulation at different contrasts with the above QPN parameters as input.
The subtracted contrast is slightly below the true contrast,
and is accounted for the offset which is typically less than 0.02.
With this approach,
we can determine the contrasts for each ensemble independent of the offset phase, ellipse fitting and bias error.

\renewcommand{\thefigure}{5}
\begin{figure}[!ht]
    \centering
    \includegraphics[width=0.95\textwidth]{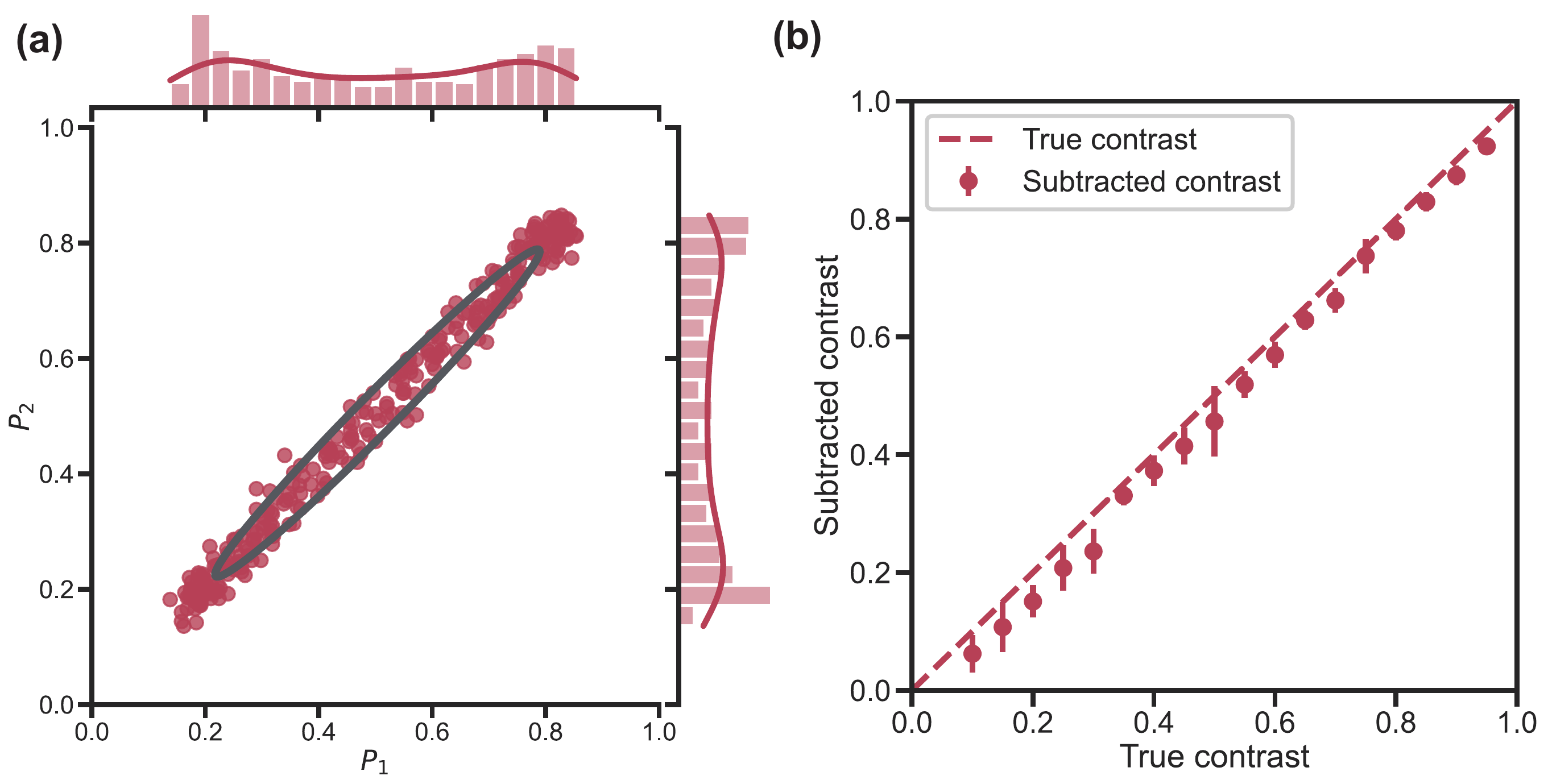}
    \caption{(a) Parametric and histogram plots showing contrast subtraction for simulated data at 0 rad offset phase with a known contrast of 0.65. 
    (b) Monte-Carlo simulation with an offset phase at 0 rad, known contrasts and QPN (500 atoms, 100 runs) as input parameters. 
    The dashed line represents the case when the subtracted contrast equals to the true contrast,
    indicating that the subtracted contrast is slightly underestimated.
    The offset will then be taken account into the subtracted contrast.
    }
    \label{figS.ContrastMC}
\end{figure}

\subsection{Loading dual isotopes into the lattice}

The experimental sequence for loading dual isotopes into the lattice is shown in Fig~S~\ref{figS.LoadingIsotopes}.
We first load ${}^{87}$Sr into the lattice,
and move the ensemble 1 cm away from the lattice center.
Unlike loading multiple ensembles of the same isotope,
here we must perform a second round of cooling in the 461-nm and 689-nm MOTs to address the second Bosonic isotope (${}^{88}$Sr, ${}^{86}$Sr or ${}^{84}$Sr) due to the isotope shifts.
This requires shifting the frequencies of the 461-nm lasers,
including 2D-MOT, Zeeman slower, 3D-MOT and probe lasers,
by as much as 270 MHz.
The frequency gap is bridged by double-passing the master 461-nm laser through two AOMs~\cite{stellmer_degenerate_2014} operating at 350 MHz with a bandwidth of about 150 MHz,
and the master laser is subsequently used to injection lock three laser diodes which are sent to the experiment table.
A simultaneous frequency tuning of up to 270 MHz within 100 ms can be achieved while maintaining the injection locking.
To efficiently cycle the 461-nm MOT for all 4 isotopes,
the two repumping lasers,
679 nm ($\ket{^3P_0}\leftrightarrow\ket{^3S_1}$) and 707 nm ($\ket{^3P_2}\leftrightarrow\ket{^3S_1}$),
are frequency modulated at 1 kHz with 1 GHz and 3 GHz amplitudes,
respectively.
For the 689-nm lasers,
the frequencies need to be shifted by about 1.5 GHz.
This is done by jumping the radio-frequency that is used to reference the optical offset phase lock. (We thank Vescent Photonics for offering us a discount on the Offset Phase Lock Servo D2-135 used to accomplish this.)
To avoid heating the ${}^{87}$Sr samples out of the lattice during the loading of the second isotope,
${}^{87}$Sr is coherently transferred and shelved in the $\ket{^3P_0}$ state via a $\pi$-pulse,
and the 679-nm repump laser is disabled during the 461-nm MOT loading for the second isotope.
A final lattice move brings the two isotopes back to the lattice center,
and two sets of imaging pulses are used to image both isotopes.

\renewcommand{\thefigure}{6}
\begin{figure}[!ht]
    \centering
    \includegraphics[width=0.85\textwidth]{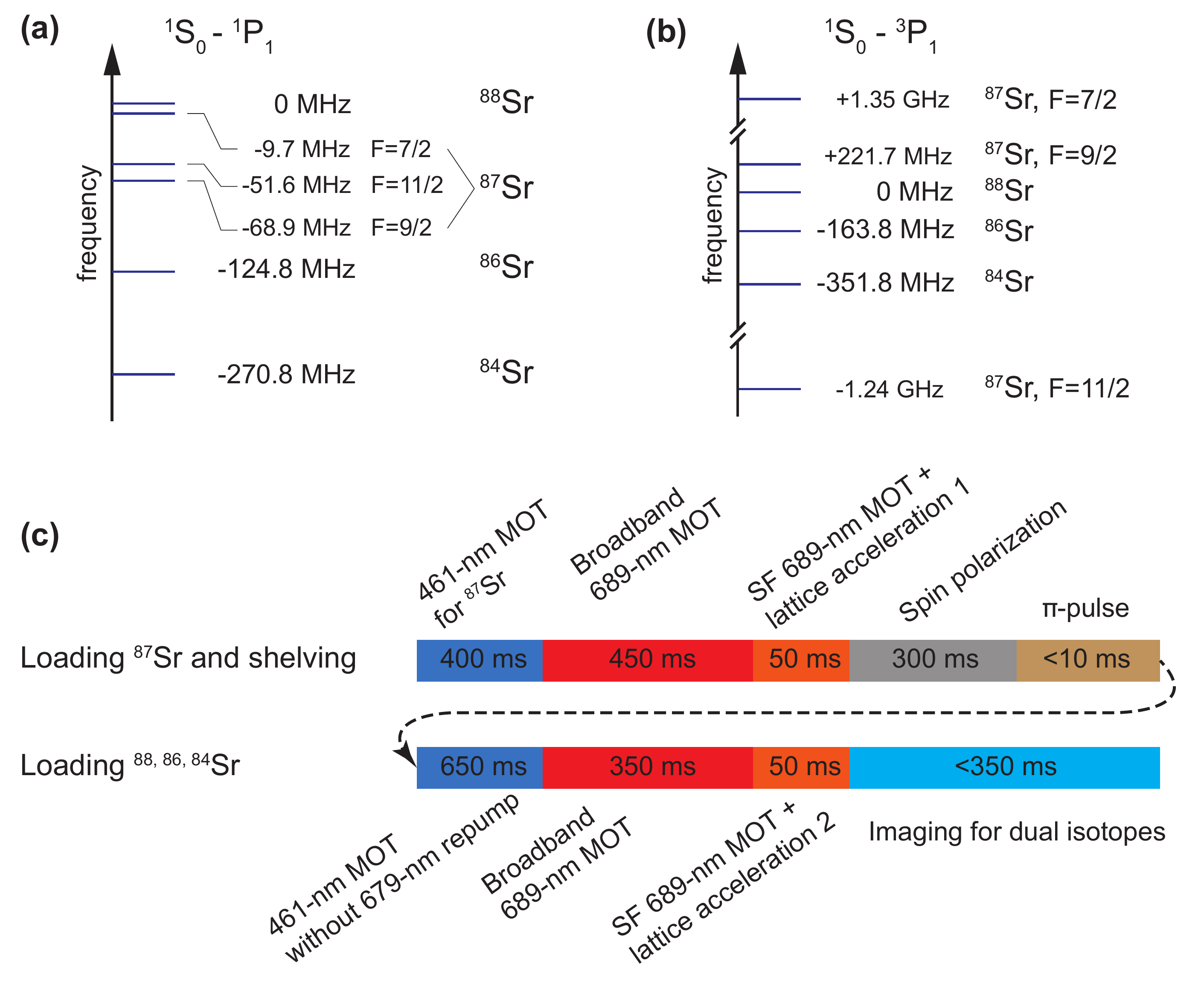}
    \caption{(a) Isotope shifts on the ${}^1S_0 \leftrightarrow {}^1P_1$ transition at 461-nm for producing the first stage MOT. The isotope shifts are relative to resonance of ${}^{88}$Sr. Note for ${}^{87}$Sr, the $F=11/2$ transition is chosen to optimize the MOT. (b) Isotope shifts on the ${}^1S_0 \leftrightarrow {}^3P_1$ transition at 461-nm for making the second stage MOT. For ${}^{87}$Sr, the $F=9/2$ and $F=11/$ transitions are in use. (c) Timing diagram for loading dual isotopes into the lattice.
    }
    \label{figS.LoadingIsotopes}
\end{figure}

\subsection{Six ensembles differential clock comparisons}

\subsubsection{Experimental sequence}

To load 6 ensembles,
we modify the loading sequence for one ensemble such that the maximal detuning is 2 MHz with 1 ms ramp time and 1.5 ms hold time,
which corresponds to a maximal velocity of 0.8 m/s and acceleration of 81 $g$.
The loading sequence is repeated for 4 times,
such that 5 subsets of the atomic ensembles can be separated from the original cloud with equal separations of 0.2 cm.
In order to reduce cross-talk due to smaller separation between each ensemble,
the imaging pulse duration is kept below 250 $\mu$s.

\renewcommand{\thefigure}{7}
\begin{figure}[!ht]
    \centering
    \includegraphics[width=0.85\textwidth]{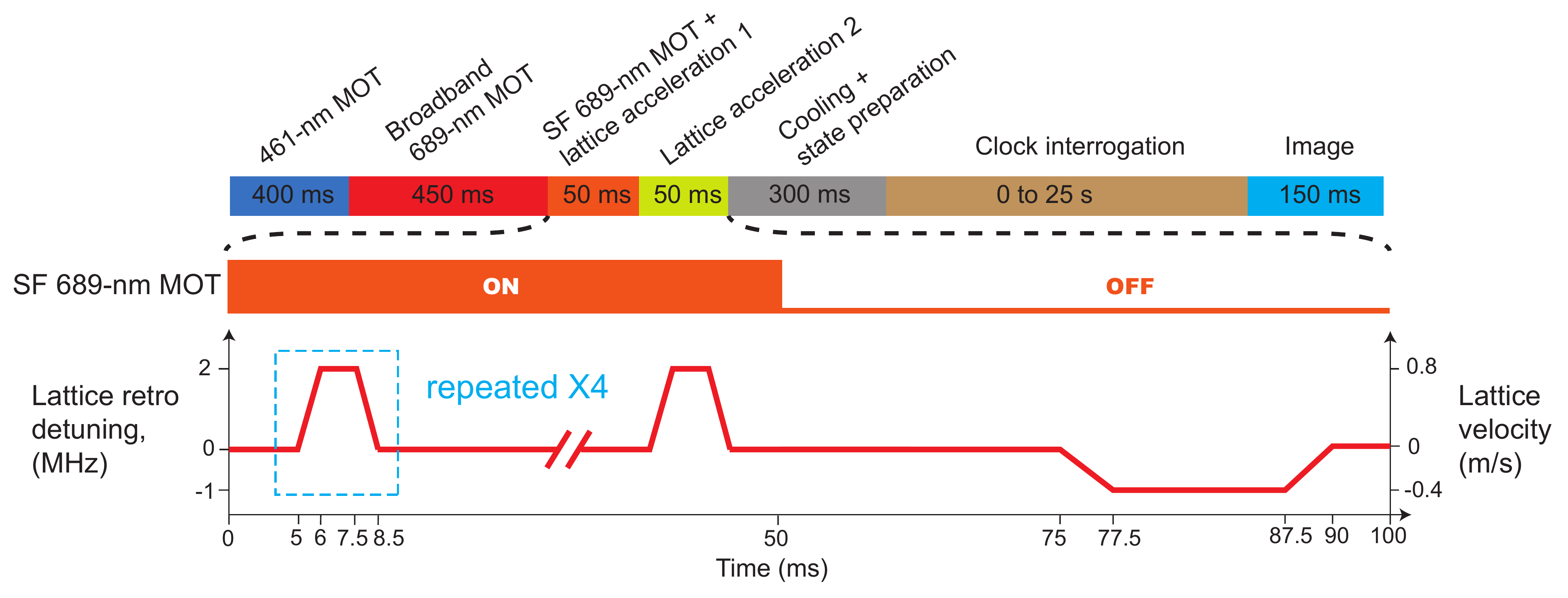}
    \caption{Timing diagram for loading 6 ensembles into the lattice.
    }
    \label{figS.6ensembles_timing}
\end{figure}

\subsubsection{Determination of differential frequencies}

To determine the differential frequencies between each clock comparison,
we run at different Ramsey times ranging from 2.5 to 8.5 s,
and the differential frequency can be mapped out through the accumulated phase evolution,
which follows
\begin{equation}
    \phi_{\textrm{acc}, ij} (T_R) = 2 \pi \delta f_{ij} T_R,
\end{equation}
where $T_R$ is the Ramsey interrogation time,
and $\delta f_{ij}  = f_j - f_i$ is the differential frequency between ensemble $i$ and ensemble $j$.
Note that $\delta f_{ij} = - \delta f_{ji}$.
While this is not entirely correct due to the bias error from ellipse fitting,
we perform Monte-Carlo simulations for each resulting pairwise ellipse using contrasts and atom numbers subtracted from each ensemble,
and correct for the bias error before mapping out the differential frequency.

\subsubsection{Numbers of unique combinations in ``Closed-loop'' self-consistency check}

We check the self-consistency of the 15 pairwise comparisons in the clock network (see Fig.4a in main text) by plotting the sum of differential frequencies within a ``closed-loop'' which has 3 or more ``clocks''.
Since the sum frequencies after clockwise and anti-clockwise rotations are equivalent,
the number of ways to arrange $n$ ``clocks'' in a loop is $(n-1)!/2$.
For a loop that has 6 ``clocks'',
for example, $(1, 2, 3, 4, 5, 6)$,
the sum frequency can be calculated as
\begin{equation}
    f = \delta f_{21}+ \delta f_{32}+ \delta f_{43}+ \delta f_{54}+ \delta f_{65} - \delta f_{61}.
\end{equation}
Note that this is equivalent to the sum frequencies of $(2, 3, 4, 5, 6, 1)$, of $(3, 4, 5, 6, 1, 2)$, of $(4, 5, 6, 1, 2, 3)$, of $(5, 6, 1, 2, 3, 4)$ and of $(6, 1, 2, 3, 4, 5, 6)$.
Therefore,
we have $5!/2=60$ unique combinations for loops of 6 ``clocks''.
For loops of 5 ``clocks'',
first there are ${}^6C_5$,
which is 6 choose 5,
combinations to choose 5 ensembles,
and $4!/2=12$ ways to arrange the ``clocks'',
therefore a total number of 72 combinations.
Similarly,
there are ${}^6C_4\times 3!/2=  45$     combinations for loops of 4 ``clocks'',
and ${}^6C_3\times 2!/2 = 20$ combinations for loops of 3 ``clocks''.
Finally,
this gives $60+72+45+20=197$ unique combinations for simultaneous clock comparisons of 6 ensembles.

\subsection{Differential density shift evaluations}

\subsubsection{Sample preparation and calibration of camera gradient}

The differential density shift can be evaluated by varying the atom number difference between symmetrically ensemble pairs.
This is accomplished by first balancing the lattice trap depths and radial profile at each ensemble by walking the focal lenses of the incoming and retro-reflecting lattice laser beams, and verified by motional sideband spectroscopy as discussed in Section A above.
The loading times into each ensemble are then varied from 0.5 ms to 20 ms to introduce imbalanced atom numbers in each ensemble,
which typically ranges from -2000 to +2000 atom number differences,
yielding a sufficiently large lever arm for differential density shifts at high $10^{-18}$ level that can be easily resolved.
The motional sideband spectrum is re-taken to ensure the temperatures of the two ensembles remain balanced after in-lattice cooling for each loading sequence.

To calibrate the camera gradient along the lattice during clock read-out,
which mainly arises from the spatial inhomogeneity of fluorescence and the imaging beam intensity gradient,
an ensemble of atoms is moved at a constant velocity of 1.5 m/s and imaged along the lattice over 500 camera pixels,
which corresponds to a distance of approximately 1.5 cm,
within less than 100 ms.
As this time scale is much smaller than the atom lifetime ($>20$ s),
the atom loss is negligible,
and a Gaussian fit to the trace of the image gives the amplitude and center (or pixel index number) of the cloud.
The average of 10 measurements,
each consisting of 100 images with randomized cloud centers spread over 500 pixels,
are taken to map out the imaging gradient.
A Savizky-Golay filter is then applied to smooth out the normalized imaging efficiency curve,
which is used for post-calibration of the camera images taken in the experiment.
A typical calibration curve with 10 averages is shown in Fig~S\ref{figS.camera_gradient}.

\renewcommand{\thefigure}{8}
\begin{figure}[!ht]
    \centering
    \includegraphics[width=0.85\textwidth]{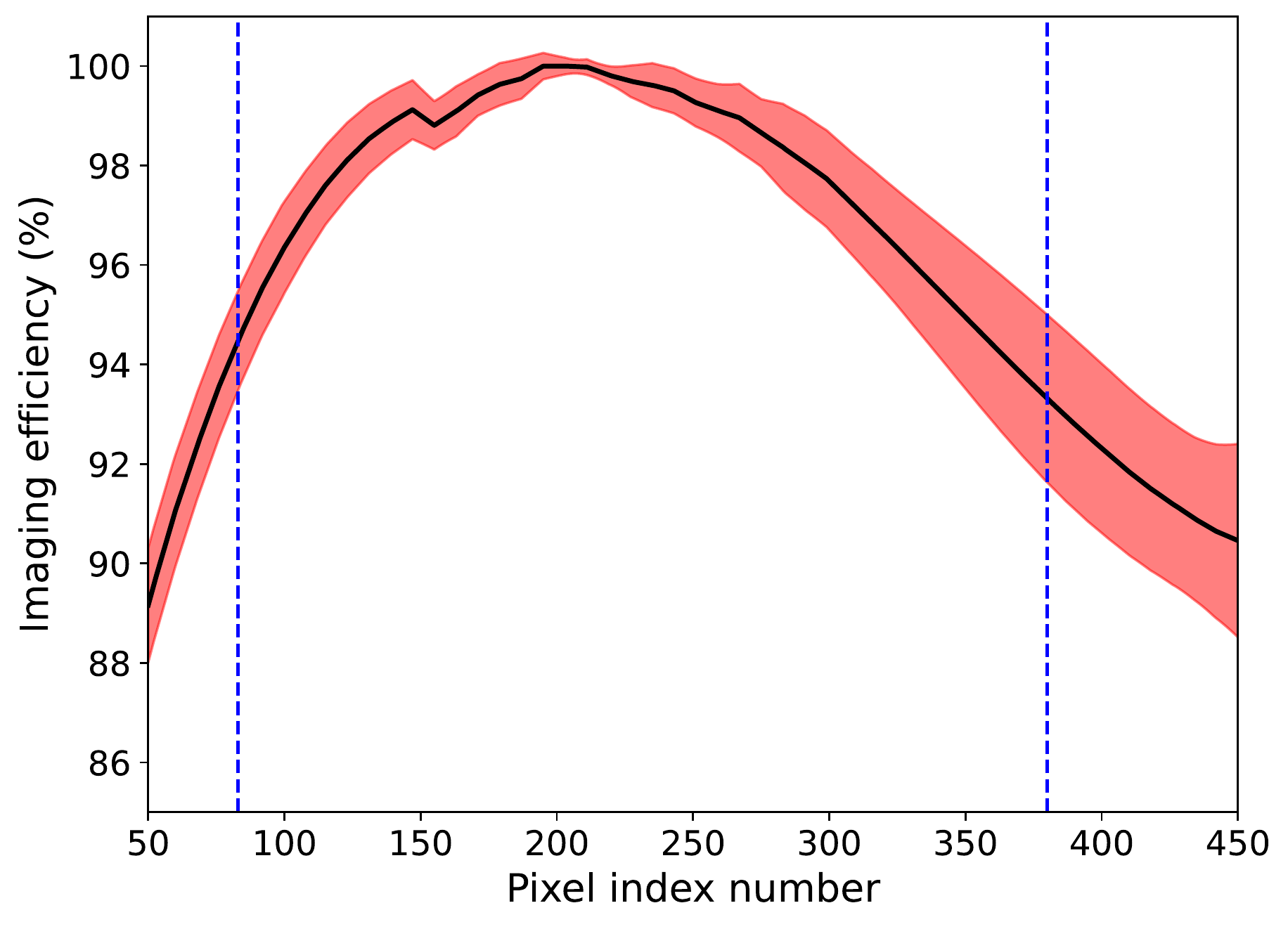}
    \caption{Calibration of fluorescence imaging gradient along the lattice. The black line is the averaged normalized imaging efficiency, the red area indicates the corresponding $1\sigma$ standard error.
    The blue dashed lines represent the pixel indices (along the gravity, $\hat{z}$) at which two ensembles are separated by 1 cm.
    }
    \label{figS.camera_gradient}
\end{figure}

\subsubsection{Differential density shift and its trap depth scaling}

Since the atomic density within an ensemble scales linearly with the atom number,
with the lattice trap depths and atomic temperatures balanced,
a change in relative atom number,
or the atom number difference between ensemble pairs,
results in a differential density shift.
To characterize this shift,
we perform a ``lock-in'' type measurement,
in which we interleave between ``low'' and ``high'' atom number differences,
i.e., $\delta N_{\textrm{low}}$ and $\delta N_{\textrm{high}}$,
which creates two ellipses with differential phase $\phi_{\textrm{low}}$ and $\phi_{\textrm{high}}$.
Each measurement yields a relative phase shift $\Delta\phi = \phi_{\textrm{high}} - \phi_{\textrm{low}}$ and relative change in atom number difference $\Delta N = \delta N_{\textrm{high}} - \delta N_{\textrm{low}}$.
This is iterated over several differential densities $\Delta N$ and averaged below $1\times 10^{-18}$ for each measurement,
as shown in Extended Data Fig.4 (a).
A linear function $f = a ~ \Delta N + b$ is applied to fit the data,
and the slope $a$ is subtracted as the density shift coefficient at 100 atom number difference.

To quantify the scaling of the differential density shift with trap depth,
the above measurement is repeated over different lattice trap depths,
and the fitted slopes are plotted as a function of trap depth, 
see Extended Data Fig.4 (b).
The data is then fitted to the model
\begin{equation}
    \alpha U^{5/4} + \beta,
\end{equation}
where $U$ is the lattice trap depth,
and $\alpha$ and $\beta$ are fit parameters.
The good agreement between the data and $U^{5/4}$ scaling implies the radial and axial trap frequencies in the lattice scale with trap depth as expected for a
thermal gas.

\printbibliography[keyword={SM},title={${}$}]

\end{document}